\documentclass[lettersize,journal]{IEEEtran}
\usepackage{amsmath,amsfonts}
\usepackage{algorithmic}
\usepackage{algorithm}
\usepackage{array}
\usepackage{textcomp}
\usepackage{stfloats}
\usepackage{url}
\usepackage{verbatim}
\usepackage{graphicx}
\usepackage{caption}
\usepackage{subcaption}
\usepackage{cite}
\usepackage{color}
\usepackage{epstopdf}
 \newcommand{\mbf}[1]{\mathbf{#1}}

\usepackage{color}  
\usepackage{mathrsfs,mathtools}
\usepackage{amssymb} 
\usepackage{enumerate}
\usepackage{bm}

\newcommand{\qed}{\hfill{$\blacksquare$}}  
\newcommand{\fin}{\hfill{$\square$}}   
\newcommand{\EL}{\mathcal{L}}
\newcommand{\N}{\mathbb{N}}
\newcommand{\R}{\mathbb{R}}
\newcommand{\w}{\bm{\lambda}}

\newcommand{\smallmtx}{\setlength\arraycolsep{1pt}
	\def\arraystretch{1}}
\newcommand{\espp}[1]{\mbox{\bf E}\left\{#1\right\}}
\newcommand{\norm}[1]{\|#1\|}

\newtheorem{example}{Example}
\newtheorem{coro}{Corollary}
\newtheorem{remark}{Remark}
\newtheorem{theorem}{Theorem}

\newtheorem{assu}{Assumption}
\newtheorem{defn}{Definition}

\begin{document}


\title{Stochastic $\EL_p$ string stability analysis in predecessor-following platoons under packet losses}

\author{Alejandro I. Maass, Francisco J. Vargas, Andr\'es A. Peters, and Juan I. Yuz
\thanks{This paper was supported by the Chilean National Agency for Research and Development (ANID) through FONDECYT Postdoctoral Grant 3230056, FONDECYT Iniciación Grant 11221365, FONDECYT Regular Grant 1241813, and FONDECYT 1230623.}
 \thanks{A.I.~Maass is with the Department of Electrical Engineering, Pontificia Universidad Católica de Chile,  Santiago, 7820436, Chile
(e-mail: alejandro.maass@uc.cl).}
 \thanks{F.J.~Vargas and J.I.~Yuz are with the Electronic Engineering Department, Universidad Técnica Federico Santa María, 2390123, Valparaíso, Chile (emails: francisco.vargasp@usm.cl, juan.yuz@usm.cl).}
\thanks{A.A.~Peters is with the Faculty of Engineering and Sciences, Universidad Adolfo Ibáñez, Peñalol\'en, 7941169, Santiago, Chile (email: andres.peters@uai.cl).}
 }

\markboth{}%
{Shell \MakeLowercase{\textit{et al.}}: A Sample Article Using IEEEtran.cls for IEEE Journals}


\maketitle

\begin{abstract}
In this paper, we study (homogeneous) predecessor-following platoons in which the vehicle-to-vehicle (V2V) communications are affected by random packet losses. We model the overall platoon as a stochastic hybrid system and analyse its string stability via a small-gain approach. For nonlinear platoons, we illustrate how the different elements of the platoon have an impact on string stability, such as platoon topology and vehicle scheduling. For linear time-invariant platoons, we provide an explicit string stability condition that illustrates the interplay between the channel success probability, transmission rate, and time headway constant. Lastly, we illustrate our results by numerical simulations.
\end{abstract}

\begin{IEEEkeywords}
Vehicle platooning, Stochastic string stability, Packet dropouts.
\end{IEEEkeywords}

\section{Introduction}
\IEEEPARstart{V}{ehicle} platooning achieved by cooperative adaptive cruise control (CACC) techniques is believed to be a key tool for reducing the impact of increasing vehicle deployment in modern road networks \cite{muratori2017potentials}. This belief is supported by the improvements and new developments in wireless communication technologies that will enable such automated cooperative driving techniques \cite{wang2019survey}. Two crucial aspects of such technologies are: the increased complexity as the relevant multi-agent systems (MAS) grow in size, and the inherent stochastic phenomena that arise with the use of wireless communications. The former is represented, for example, by scalability issues such as ``string instability" \cite{sepahe04}, that is, the amplification of disturbances as they propagate along an interconnection of systems. To deal with scalability issues under network-induced communication constraints, it is advantageous to analyse the platoon using networked control systems theory  \cite{li2017}. In this context, it is crucial to consider not only the platoon parameters (such as vehicle dynamics, topology, and time headway spacing policy) but also to investigate their interaction with the network parameters (such as probability of packet loss, scheduling, and transmission rate).

 Although string stability for the deterministic case has been deeply studied for over fifty years \cite{lveath66,swahed96,middleton2010,gunter2020commercially}
the notion of stochastic string stability remains sparsely explored, see \cite{socha2004,rybarska2007string} for the earliest technical approaches to stochastic string stability of interconnected string systems. As discussed in a recent survey on platooning and string stability \cite{feng2019string}, general sufficient conditions on the design parameters of CACC schemes for string stability in a stochastic setting are still not available, but more precisely, there is a lack of appropriate mathematical tools for answering this and related questions. For linear platoons under lossy channels, most works are simulation-based studies on the effects of packet losses in stochastic string stability, see e.g. \cite{vargas2018,gordon2021comparison,Lei2011,van2017robust,acciani2018obs,villenas2023kalman}. 
In this particular context, more comprehensive research has been conducted on stochastic string stability in \cite{acciani2021stochastic,li2019string,zhao2020,rezaee2024cooperative}. It is worth noting that these studies offer different approaches and definitions for analysing and understanding stochastic string stability.
For additive noise channels and linear platoons, conditions for the so-called \emph{mean square string stability} and \emph{$\mathcal{L}_p$-mean $\mathcal{L}_q$-variance  string stability} are presented in \cite{vargor24} (see also \cite{gordon2020platoon}).
We highlight that the study of stochastic string stability for non-linear platoons remains much less developed, see e.g. \cite{shi2015stability,socha2018exponential} for classes of input-affine non-linear systems.
In a deterministic setting, string stability of non-linear platoons has been studied in the literature by extending the classical (input-output) $\mathcal{L}_p$ stability property of non-linear systems to include the scalability of the platoon, leading to the concept of \emph{$\mathcal{L}_p$ string stability} \cite{ploeg2013lp,Besselink2017,monteil2018,monteil2019string,feng2020tube}.
The topic of extending the $\mathcal{L}_p$ string stability definition to stochastic settings has attracted a great deal of attention lately, as evidenced by \cite{acciani2021stochastic,li2019string,zhao2020}, motivated by recent surveys reporting  poor development of the stochastic theory for scalability issues in platooning applications \cite{feng2019string}. 

In this work we are particularly interested in the stochastic phenomena introduced by packet losses in vehicle-to-vehicle (V2V) communications. In this context, 
the authors of \cite{teo2003decentralized} study the effect of random momentary lead vehicle state dropouts, concluding that ``leader-to-formation stability" is retained if the lead vehicle maintains its speed, while the followers estimate the lead vehicle's state using dead reckoning. On the other hand, \cite{Lei2011,vargas2018} established through simulation studies that packet loss, even at the nearest neighbour communication level, has a negative effect on the string stability properties of platooning schemes. The simulation study in \cite{gordon2021comparison} illustrates that some data-loss compensation strategies provide better results in terms of performance, while some others present
improvements for string stabilisation.
The work \cite{van2017robust} discusses, with simulations and experimental results, the benefits of using MPC for dealing with packet dropouts, using a buffer at every vehicle to store the model-based predictions of the desired accelerations received from the nearest front neighbour. 
The use of an observer to constantly deal with packet-loss in a discrete-time setting CACC is reported in \cite{acciani2018obs}, where agents receive the control input of the nearest predecessor wirelessly. CACC design with lossy communications over ``average dynamics" was recently proposed in \cite{acciani2021stochastic}, where string stability was ensured using $\mathcal{H}_\infty$ control tools. In \cite{li2019string} an $\mathcal{L}_2$ stochastic string stability definition is presented and then used to study an event-triggered platooning scheme with unreliable communications. In \cite{zhao2020}, Markov jump linear systems theory was adopted to cast a minimisation problem, whose feasibility ensures that a control design achieves string stability for a platooning scheme with random packet drops.
More recently, \cite{rezaee2024cooperative} proposes a specific adaptive control strategy which ensures \emph{almost surely $\mathcal{L}_{\infty}$ string stability} under packet dropouts.
Other related works study mean square stability of the underlying platoon  \cite{gordon2023mean,tang2018consensus,elahi2022distributed}, however, string stability in the corresponding stochastic setting is not analytically studied. Lastly, \cite{villenas2023kalman} provides an estimation scheme based on an intermittent Kalman filter, and string stability is only analysed via simulations.

With the exception of \cite{acciani2021stochastic,li2019string,zhao2020,rezaee2024cooperative}, most of the works listed above primarily focus on analysing the impact of packet losses on stochastic string stability through simulations or internal stability (excluding string stability). In this paper, our objective is to go beyond these existing works by introducing a rigorous formalism that enables the analysis of string stability in predecessor-following platoons affected by data losses. 
%
We build upon the notion of $\mathcal{L}_p$ string stability, as proposed for example in \cite{ploeg2013lp,Besselink2017,monteil2018,monteil2019string,feng2020tube}, and adapt it to deal with the proposed stochastic setting.
Specifically, this paper presents the following key contributions:

    $(i)$ In contrast to prior works like \cite{acciani2021stochastic,li2019string,zhao2020,rezaee2024cooperative}, which concentrate  on either purely discrete-time or purely continuous-time platoon models, our study addresses a substantially broader scenario by employing stochastic hybrid systems (SHSs). Specifically,  both continuous and discrete time dynamics observed in wireless platoons are included in the model, thereby integrating inter-sample behaviour for a more comprehensive analysis. Furthermore, this formalism enables the inclusion of additional features such as  vehicle scheduling and stochastic transmission instants, which was not considered in \cite{acciani2021stochastic,li2019string,zhao2020,rezaee2024cooperative}.

    $(ii)$ For general non-linear cascaded systems, we offer a comprehensive interpretation of the interplay between factors such as transmission rate, scheduling, topology, and more, with respect to string stability.  Related literature mostly concentrates on input-affine non-linear systems, see e.g. \cite{shi2015stability,socha2018exponential}.

    $(iii)$ For the linear case, we provide explicit string stability conditions, in contrast to existing works that primarily rely on simulation-based studies  \cite{vargas2018, gordon2021comparison, Lei2011, van2017robust, acciani2018obs, villenas2023kalman}.
     Specifically, we establish closed-form relationships between crucial vehicle and network parameters such as the time headway constant $h$, the probability of successful transmission $p$, and the transmission rate %
     We develop these results for two relevant scenarios of data loss. Specifically, when solely the desired acceleration of the predecessor is subject to losses, and also in situations where the predecessor's position and velocity may also be lost.
     A similar relationship between $h$ and $p$ was previously identified through simulations in \cite{vargas2018}, however, this paper presents explicit theoretical conditions. It is also worth noting that \cite{zhao2020} also obtained  theoretical conditions, but regarding the internal stability of the platoon rather than string stability.


\textbf{Notation:}
Let $\mathbb{N}\coloneqq\{1,2,3,\dots\}$, $\N_0\coloneqq \N\cup\{0\}$, $\R^n$ be the set of all real vectors with $n\in\N$,  $\R^{m\times n}$ be the set of all real matrices of dimensions $m\times n$, $\R_{>0}\coloneqq (0,\infty)$ and $\R_{\geq 0}\coloneqq [0,\infty)$.
For any $x\in\R^n$ and $y \in \R^m$, we define $(x, y)\coloneqq [x^\top\ y^\top]^\top \in \R^{n+m}$.
For $x \in\R^n$, $|x|$ denotes the standard Euclidean norm, and also the induced 2-norm for a real matrix.
Given a (Lebesgue) measurable function $f : \R \to \R^n$, $\norm{f}_{\EL_p}
\coloneqq \left( \int_{\R} |f(s)|^p \mathrm{d}s \right)^{1/p}$, for $p \in \N$, and $\norm{f}_{\EL_{\infty}}\coloneqq 
\mbox{ess} \sup_{t\in\R} |f(t)|$.
Given an interval $[a,b]\subset\R$, 
$\norm{f}_{\EL_p[a,b]}
\coloneqq \big( \int_{a}^{b} |f(s)|^p \mathrm{d}s \big)^{1/p}$
and 
$\norm{f}_{\EL_{\infty}[a,b]}
\coloneqq \mbox{ess} \sup_{t\in[a,b]}
|f(t)|$. We say that $f\in\EL_p$ for $p \in \N \cup \{+\infty\}$ whenever
$\norm{f}_{\EL_p}< \infty$.
Given $t \in \R$ and a piecewise continuous function $f : \R \to
\R^n$, we define $f(t^+) \coloneqq \lim_{s\to t,s>t}
f(s)$.
The underlying complete probability space is 
$(\Omega, \mathcal{F}, \mathbb{P})$, with $\Omega$ the sample space, $\mathcal{F}$ the $\sigma$-algebra, and $\mathbb{P}\{\cdot\}$ the probability measure.
The expectation operator is denoted by $\mathbf{E}\{\cdot\}$. For a measurable function $g : \Omega \times \R \to \R^n$, we say
that $g \in \mathcal{L}_p^e$
 whenever $\mathbf{E}\{\norm{g}_{\EL_p}\} < \infty$. By \emph{i.i.d.} we mean independent
and identically distributed.

\section{Problem setting}\label{sec:setup}
\begin{figure}
	\begin{center}
		\includegraphics[scale=0.9]{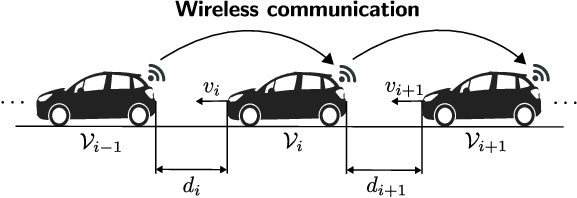}       
		\caption{Platoon configuration.}  
		\label{fig:platoon}                      \end{center}                \end{figure}
\subsection{Platoon description}
Consider a platoon of $N\geq 2$ identical vehicles, illustrated schematically in Figure \ref{fig:platoon}, where $d_i$ represents the distance between vehicle $\mathcal{V}_i$ and its preceding vehicle $\mathcal{V}_{i-1}$, and $v_i$ the velocity of vehicle $\mathcal{V}_i$.
For each vehicle, the primary goal is to track its preceding vehicle while maintaining a desired distance, denoted as $r_i$.
We adopt a constant time-headway policy, whose objective is to maintain a consistent spacing between vehicles based on their velocities.
This is formulated as $r_i(t) = \varepsilon_i + hv_i(t)$, for $i\in\mathcal{N}$, $\mathcal{N}\coloneqq \{1,\dots,N\}$. , where $h\geq 0$ is referred to as the time headway constant, and $\varepsilon_i$ is the standstill distance. This policy aims at decreasing the chance of a collision by increasing the \emph{time gap} between vehicles \cite{swaroop2001}; and it 
is known to improve string stability \cite{rajamani2002semi} and safety \cite{ioannou1993autonomous} when using only nearest neighbour information. 
In this context, each vehicle sends a reference acceleration over the corresponding Dedicated Short-Range Communication (DSRC) \cite{nguyen2017comparison}, which can be subject to data loss. 
We assume the platoon is homogeneous in the sense that every vehicle uses the same controller, has the same spacing policy, the same model, and the same probability of success for the corresponding DSRC channel. 

Let the spacing error be defined by $\xi_i(t) \coloneqq d_i(t) - r_i(t)=[s_{i-1}(t)-s_i(t)-L_i]-[r_i + hv_i(t)]$, $i\in\mathcal{N}$, where $s_i$ 
and $L_i$ denote the position and length of vehicle $\mathcal{V}_i$, respectively.
The first vehicle in the platoon $\mathcal{V}_1$ follows a \emph{virtual reference vehicle} denoted by $\mathcal{V}_0$. Then, $\xi_1$ corresponds to the spacing error between the leader and this virtual reference.

The following vehicle model is adopted as the basis for control design, see e.g. \cite{ploeg2013lp,dolk2017event,li2019string},
\begin{align}\label{eq:vehicle}
    \mathcal{V}_i:\begin{bmatrix}
        \dot{s}_i(t) \\
        \dot{v}_i(t) \\
        \dot{a}_i(t)
    \end{bmatrix}
    = 
    \begin{bmatrix}
         v_i(t) \\
        a_i(t) \\
        -\frac{1}{\tau}a_i(t) + \frac{1}{\tau}u_i(t)
    \end{bmatrix},\ i\in\mathcal{N}\cup \{0\},
\end{align}
where $a_i$ denotes the acceleration of vehicle $\mathcal{V}_i$, $u_i$ the external input (reference acceleration), and $\tau$ the characteristic time constant representing drive-line dynamics. 

\begin{figure*}
	\begin{center}
		\includegraphics[scale=0.9]{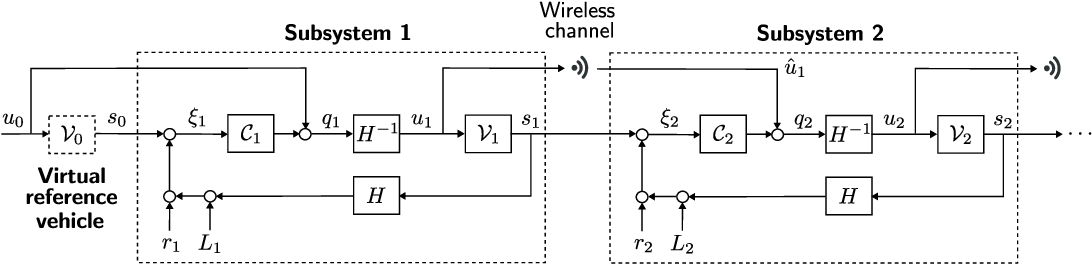}       
		\caption{Block diagram of the CACC scheme.}  
		\label{fig:block_diagram}                      \end{center}
  \end{figure*}

In CACC schemes, the control law is typically designed based on the spacing error $\xi_i$ and a feedforward component being the direct feedthrough of  $u_{i-1}$. 
Inspired by \cite{ploeg2011design,ploeg2013lp}, 
we consider the CACC scheme in Figure \ref{fig:block_diagram},
where the control law is denoted by $q_i$, and it is filtered by $\dot{u}_i = -\frac{1}{h}u_i + \frac{1}{h}q_i$ before going into the vehicle drive-line in \eqref{eq:vehicle}. 
This filter is given by $H(s)=hs+1$.
The control law $q_i$ has a feedback component generated by the PD controller $\mathcal{C}_i(s)= k_p + k_d s$, and feedforward component $\hat{u}_{i-1}$. Formally, the control law takes the form
\begin{align}\label{eq:controller}
    q_i(t) = k_p \xi_i(t) + k_d \dot{\xi}_i(t) + \hat{u}_{i-1}(t),\ i\in\mathcal{N},
\end{align}
with controller gains $k_p$ and $k_d$ to be designed.
The feedback component of the controller often utilises measurements from a forward-looking radar. As such, we assume that each vehicle has the ability to measure the relative distance and relative speed to the preceding vehicle (via the radar sensor), as well as its own absolute speed and acceleration. However, obtaining relative accelerations through local onboard sensors is challenging, and therefore, they are commonly acquired through wireless communications, see e.g. \cite{acciani2021stochastic,dolk2017event}.
We thus use $\hat{u}_{i-1}$ to denote the signal $u_{i-1}$ but received through the DSRC channel from vehicle $i-1$.
Because of the packet-based nature of the communication channel and the existence of packet dropouts, it is generally the case that $\hat{u}_{i-1}(t)\neq u_{i-1}(t)$ for $t\in\mathbb{R}_{\geq 0}$. 
Note that $\hat{u}_0(t) = u_0(t)$ for all $t\in\R_{\geq 0}$ as the first vehicle follows a virtual reference (i.e. no network imperfections). We will describe the dynamics of the communication channel, and thus $\hat{u}_{i-1}$, in detail in the following section.

\begin{remark}\label{rem:local-sensors}
We highlight that the results and methodology developed in this paper can be used for other platoon settings such as where not only $u_{i-1}$ is transmitted wirelessly, but also the GPS-measured position $s_{i-1}$ and speed $v_{i-1}$. This way, local sensors can be fully dedicated to obstacle avoidance or as a redundancy mechanism, see e.g. \cite{li2019string}.  We illustrate this fact at the end of Section 
\ref{sec:string-stability} (cf. Theorem \ref{theo:SS-explicit-new}).\fin 
\end{remark}

\subsection{Main objectives}
Typically, a well-designed CACC should fulfil two main objectives. The first objective is vehicle following, which involves regulating spacing errors and is known as \emph{individual vehicle stability}. The second objective is to avoid the amplification of disturbances that propagate along the vehicle platoon, known as \emph{string stability}, as mentioned in the introduction. 
To achieve these objectives in a stochastic setting with packet-based lossy communications between vehicles, we will follow an \emph{emulation approach} \cite{wabebu01}.
That is, we first design the controller \eqref{eq:controller} when network issues are not considered, i.e., when $\hat{u}_{i-1} = u_{i-1}$ in \eqref{eq:controller}. 
An advantage of this approach is that existing design techniques can be used for the platoon. In fact,
the network-free design that ensures both individual vehicle stability and string stability has been well studied in the literature. Particularly, as shown in \cite{ploeg2013lp}, the required platoon stability properties are satisfied for any $h,k_d,k_p>0$ for which $k_d>k_p\tau$. 
At the second stage of emulation, we implement the designed network-free controller over the wireless platoon and study the effect that the network imperfections have on string stability. Specifically, our goal is to provide sufficient conditions on the network and vehicle parameters that ensure the desired stability properties of the network-free system are preserved  despite the impact of packet-based communication and packet losses.

\subsection{String stability notion}
Since the transmitted signals $\hat{u}_{i-1}$ may be stochastic given the network effects, we need an appropriate notion of string stability for our setting. We present a definition motivated by the (input-output) \emph{$\EL_p$ string stability} property proposed in \cite{ploeg2013lp}, which has been extensively used in the literature, see e.g. \cite{monteil2018,monteil2019string}.
As our setting is stochastic, we introduce the notion of \emph{$\EL_p$ string stability in expectation} as per the definition below.
Let $x_i\coloneqq (\xi_i,v_i,a_i,u_i)$, $i\in\mathcal{N}$, $w\coloneqq (v_0,u_0)$, and $\mbf{x}\coloneqq (x_1,\dots,x_N)$.
\begin{defn}\label{def:Lp-MSS}
Let $p\in\N\cup\{+\infty\}$. We say that the wireless platoon in Figure \ref{fig:block_diagram} is \emph{$\EL_p$ string stable in expectation} if there exist non-negative constants $K$ and $\gamma$ such that, for any $\mbf{x}(0)\in\R^{n_x}$ and $w\in\EL_p$,
\begin{align}\label{eq:Lp-MSS}
	\espp{\norm{x_i}_{\EL_p[0,t]}} \leq K|\mbf{x}(0)| + \gamma \norm{w}_{\EL_p[0,t]},\ \forall t\geq 0,
\end{align}
for all $i\in\mathcal{N}$ and all platoon lengths $N\geq 2$.\fin 
\end{defn} 

In contrast to the standard $\EL_p$ stability, 
we emphasise that \eqref{eq:Lp-MSS} must hold for any platoon length $N$ to guarantee  \emph{string} stability. That is, the constants $K$ and $\gamma$ are independent of the number of vehicles. 
This implies that a string stable system possesses a so-called scalability property, where the required stability notion is not affected by the removal or addition of vehicles \cite{feng2019string}. Whether or not a particular system satisfies \eqref{eq:Lp-MSS} will be determined by the parameters of the vehicles, local controllers, spacing policy, and more importantly within this setting, by the network parameters (and in more general cases, by the topology of the interconnected system). 
We show that these conditions depend on the probability of successful transmission, the time-headway constant, the implemented scheduling protocol, and the rate of transmission.

A notion similar  to Definition \ref{def:Lp-mean} when $p=2$ has been introduced in \cite{li2019string}, and called \emph{stochastic $\EL_2$ string stability}.
Moreover, in stochastic networked control systems (NCSs) in general, similar definitions \emph{in expectations} have been used, but without taking into account the scalability (``string") aspect. See for example the $\EL_p$ \emph{stability in expectation} property as introduced in \cite{tabnes08-tac}. 
We observe that, similar to previous research focusing on input-output $\EL_p$ properties, the disturbances within the scope of Definition \ref{def:Lp-MSS} are assumed to be both $\EL_p$-bounded and deterministic. In the context of our particular platooning configuration, the variable $w$ denotes changes in the leader's velocity and acceleration. Hence, our main objective in this paper is to explore how these changes impact the stochastic string stability of the underlying platoon.

\begin{remark}
We emphasize that the appropriate choice for the norm (i.e. the value of $p$ in \eqref{eq:Lp-MSS}) in this context depends on the underlying platooning configuration, as illustrated in \cite{vargor24}. If our concern is the $\EL_2$ norm, we focus our attention to velocity/acceleration profiles $w=(v_0,u_0)$ characterised by a bounded 2-norm, as discussed in \cite{monteil2018}. It is worth noting that other common velocity profiles converge to a constant, possibly requiring a different definition of the state $\mbf{x}$ (refer to \cite{vargor24, gordon2023mean, zhao2020, li2019string}) to ensure the  norms remain bounded. Incorporating these modifications into the framework presented in this paper, we believe that does not require  significant additional effort.\fin 
\end{remark}

\section{Inter-vehicle wireless communication}\label{sec:network}
We first describe the DSRC channel between vehicles, which, in turn, determines the dynamics of the transmitted signals $\hat{u}_{i-1}$ in \eqref{eq:controller}. The wireless network is characterised by a set of transmission instants $\mathcal{T}$, a packet loss process $\theta$, and a time-varying matrix $\Psi$ that represents the underlying scheduling protocol. Next, we proceed to define each of them.
%


\subsection{Transmission instants}
We define $\mathcal{T}\coloneqq \{t_0,t_1,t_2,\dots\}$ as the unbounded set of times at which the predecessor's information $u_{i-1}$, $i\in\mathcal{N}\setminus\{1\}$, is transmitted. In similar studies \cite{acciani2021stochastic,li2019string,zhao2020}, transmission instants have been assumed to be equidistant. As a way of contrast, we here examine a scenario that accounts for the randomness of transmission times in wireless channels.
Indeed, 
 transmission times are typically neither equidistant nor deterministic, but rather exhibit randomness due to synchronisation times, acknowledgements, waiting times, etc.
For example, for networks with carrier sense multiple access, transmissions occur randomly as devices wait for channel clearance under \emph{random back-off mechanisms}.
Consequently, we assume the following:
\begin{assu}\label{assu:transmission-instants}
	Consider a Poisson point process $r(t)$ with rate $\w\in\R_{>0}$ that satisfies $r(t)=0$ for $t\in[0,t_0)$ and $r(t)=k$ for $t\in[t_{k-1},t_k)$, where $t_k\in\mathcal{T}$, $k\in\N_0$, are defined inductively by: $t_0=\tau_0$ with $\tau_0\sim \text{Exp}(\w)$, and for each $k\in\N$, $t_k=t_{k-1}+\tau_k$, with $\tau_k\sim \text{Exp}(\w)$, where the sequence $\{\tau_k\}_{k\in\N_0}$ is i.i.d.\fin
\end{assu}
This modelling approach has also been used in NCS literature such as \cite{tabnes08-tac,maass2023state}.
The times $\{t_k\}_{k\in\N_0}$ are also called \emph{arrival times} \cite{tijms03}, $\{\tau_k\}_{k\in\N_0}$ are called \emph{inter-transmission times} (or \emph{inter-arrival times}),   $\bar{\tau}\coloneqq 1/\w$ represents the \emph{average inter-transmission time}, and $\w$ is the \emph{arrival rate}. Throughout this paper, we use the terms arrival rate and transmission rate interchangeably.
The exponential distribution that governs each $\tau_k$ describes the time between transmissions. 

\subsection{Packet losses}
The next element that characterises the dynamics of the transmitted signals $\hat{u}_{i-1}$ are packet dropouts or packet losses. 
Two events may occur at transmissions, either the vehicle $i$ receives the packet from its predecessor $i-1$ successfully (i.e. $\hat{u}_{i-1}=u_{i-1}$), or it is lost with some probability $\alpha\in[0,1]$. To model this behaviour, we let
$\{\theta(k)\}_{k\in\N}$ be an i.i.d. Bernoulli sequence such that $\theta(k)=1$ with probability $\alpha$ (\emph{probability of successful transmission}), and $\theta(k)=0$ with probability $1-\alpha$.

Due to packet losses, it is useful to define the so-called \emph{network-induced error}, which represents the error present in the information $\hat{u}_{i-1}$ available at vehicle $\mathcal{V}_i$, with respect to the information $u_{i-1}$ sent from vehicle $\mathcal{V}_{i-1}$. We define this error as
$e_{u_{i-1}}\coloneqq \hat{u}_{i-1}-u_{i-1}$, $i\in\mathcal{N}\setminus\{1\}$, noting that $e_{u_0}$ (which is zero) is excluded as the leader follows a virtual reference without network imperfections. 

We now use the network-induced error to model the dynamics of the V2V communications, and specifically, when vehicle $\mathcal{V}_{i-1}$ is scheduled to transmit $u_{i-1}$ to  vehicle $\mathcal{V}_i$ at time instant $t_k$. Then, we assume $e_{u_{i-1}}(t_k^+)=0$ (i.e. $\hat{u}_{i-1}(t_k^+)=u_{i-1}(t_k)$) only if the transmission is successful ($\theta(k)=1$). Here, $e_{u_{i-1}}(t_k^+)$ denotes the right limit of $e_{u_{i-1}}(\cdot)$ at time $t_k$. On the other hand, whenever a packet loss occurs, then we assume that the corresponding error components remain unchanged since the signal was not updated, i.e., $e_{u_{i-1}}(t_k^+)=e_{u_{i-1}}(t_k)$ when $\theta(k)=0$.
We note that compensation strategies to cope with packet loss in vehicle platooning could be potentially used (see e.g. \cite{gordon2021comparison}), but they are beyond the scope of this paper.
The above description is captured by
\begin{align}\label{eq:protocol-loss}
\mathbf{e}(t_k^+) = \underbrace{\theta(k)\Psi(k)\mathbf{e}(t_k) + (1 - \theta(k))\mathbf{e}(t_k)}_{=:\mbf{h}(k,\mbf{e}(t_k))},
\end{align}
where $\mathbf{e}\coloneqq (e_{u_1},\dots,e_{u_{N-1}})\in\R^{n_e}$, $n_e\coloneqq N-1$, and  $\Psi(k)$ is a time-varying diagonal matrix containing only 1's and 0's that models the implemented scheduling protocol. This matrix sets to zero components of $\mathbf{e}$, and thus determines which vehicle transmits and when.
Further below we provide examples of protocols and their corresponding $\Psi(k)$.

Lastly, in between two transmission events, the value of $\hat{u}_{i-1}$ is kept constant in a zero-order hold fashion. As such, we assume that $\dot{\hat{u}}_{i-1}=0$, $i\in\mathcal{N}\setminus\{1\}$, for $t\in[t_{k-1},t_{k}]$, and any $t_k\in\mathcal{T}$. This is common in multi-hop networks where communicating devices behave like routers/buffers to receive and transmit packets. 


\subsection{Scheduling protocols}
We are also interested in studying the role that vehicle scheduling has in string stability.
As per \eqref{eq:protocol-loss}, this is characterised by the time-varying matrix $\Psi$. We consider the following assumptions on the underlying class of protocols, which is standard in NCS literature, see e.g. \cite{tabnes08-tac}.
\begin{assu}\label{assu:as-UGES}
	Let $W:\N_0\times\R^{n_e}\rightarrow \R_{\geq 0}$ be given and suppose there exists a sequence of non-negative independent random variables $\{\kappa_k\}_{k\in\N_0}$, and positive real numbers $\underline{a}_W,\overline{a}_W,\bar{\kappa}\in\R_{>0}$ such that the following conditions hold for the auxiliary discrete-time system $\mbf{e}(k+1)=\textbf{h}(k,\mbf{e}(k))$:
	\begin{subequations}
		\begin{align}
		\underline{a}_W|\mbf{e}| &\leq W(k,\mbf{e}) \leq \overline{a}_W|\mbf{e}|,\label{eq:W-a1a2} \\
		W(k+1,\textbf{h}(k,\mbf{e})) &\leq \kappa_kW(k,\mbf{e}),\label{eq:W-kappa}\\
		\mathbf{E}\{\kappa_k\} &\leq \bar{\kappa} < 1, \label{eq:W-esp-kappa}
		\end{align}
	\end{subequations}
for all $k\in\N_0$ and all $\mbf{e}\in\R^{n_e}$.\fin 
\end{assu}
As defined in \cite{tabnes08-tac}, 
we say that protocols satisfying Assumption \ref{assu:as-UGES} are \emph{almost surely uniformly globally exponentially stable} (a.s. UGES) with Lyapunov function $W$. This class of protocols is the stochastic counterpart for the well-known deterministic UGES property introduced in \cite{nestee04}.
It can be shown that any UGES protocol is a.s. UGES for a non-zero probability of successful transmission, and many protocols satisfy this property. We provide two examples below.

\begin{example}[Sampled-data (SD) protocol \cite{oncu2012string}]\label{ex:SD}
This protocol samples and wirelessly transmits all vehicle data simultaneously, which in practice can be achieved by GPS clock synchronisation.
Here $\Psi(k)=0$ for all $k\in\N_0$, i.e. when there is no packet loss, the error vector $\mbf{e}(t_k^+)=0,\forall k\in\N_0$. That is, every vehicle transmits its data at every $k\in\N_0$. This is a common assumption  made in the literature on discrete-time platooning.
Let $W(k,\mbf{e})=|\mbf{e}|$, then we can see that $\underline{a}_W=\overline{a}_W=1$, and that
\begin{align*}
	W(k+1,\textbf{h}(k,\mbf{e})) &= |\theta(k)\Psi(k)\mbf{e} + (1 - \theta(k))\mbf{e}| \\
	&\leq (1-\theta(k))W(k,\mbf{e}).
	\end{align*}
	Consequently, $\kappa_k = 1 - \theta(k)$ and thus $\bar{\kappa} = 1-\alpha$. Note that the SD protocol satisfies Assumption \ref{assu:as-UGES} for any $\alpha\in(0,1]$.
\end{example}

\begin{example}[Round-robin (RR) protocol \cite{nestee04}]\label{ex:RR}
	 This protocol is employed in token ring and token bus network protocols \cite{wheels93}. Vehicle data is transmitted in a predetermined and cyclic manner. Similar to the SD protocol, it is possible to show that RR protocol is a.s. UGES with $\underline{a}_W=1$, $\overline{a}_W=\sqrt{N-1}$, and $\bar{\kappa}=1-\alpha+\alpha\sqrt{N-2/N-1}$, which is strictly smaller than 1 for any $\alpha\in(0,1]$, as shown in \cite[Proposition 5.3]{tabnes08-tac}.
\end{example}

 We emphasize that for RR protocol, both $\bar{\kappa}$ and $\overline{a}_W$ depend on the number of vehicles $N$, but for SD protocol, $\bar{\kappa}$ and $\overline{a}_W$ are independent of $N$. We illustrate further below that this fact is crucial when analysing string stability, as the protocol parameters have a direct impact on the gains associated with Definition \ref{def:Lp-MSS}.


\section{String stability: An initial treatment}\label{sec:string-stability}

In this section we provide an initial treatment on stochastic string stability, 
outlining general guidelines that highlight the role of vehicle and network parameters in attaining string stability.

\subsection{Hybrid system formulation for the platoon}\label{sec:hybrid-model}
 Hybrid systems are a very useful tool to model networked control systems, as evidenced by, e.g., \cite{tabnes08-tac,hestee06}. We thus start with a hybrid systems formulation for the platoon.
Based on the descriptions in Sections \ref{sec:setup} and \ref{sec:network}, we can write the following state-space representation of the $i$-th subsystem in Figure \ref{fig:block_diagram} during flows (i.e. $\forall t\in[t_{k-1},t_k]$),
\begin{align*}
    \dot{x}_1(t) &= A x_1(t) + B_w w(t) \\
    \dot{x}_i(t) &= A x_i(t) + B x_{i-1}(t) + B_e e_{u_{i-1}}(t) \\
    \dot{e}_{u_1}(t) &= -C_uA x_1(t) -C_u B_w w(t) \\
    \dot{e}_{u_{i-1}}(t) &= - C_uA x_{i-1}(t) + C_uB x_{i-2}(t) + C_uB_e e_{u_{i-2}}(t)
\end{align*}
for all $i\in\mathcal{N}\setminus \{1\}$, where
\begin{align*} 
    A
    &=
    \begin{bmatrix}
        0 & -1 & -h & 0 \\
        0 & 0 & 1 & 0 \\
        0 & 0 & -\frac{1}{\tau} & \frac{1}{\tau} \\
        \frac{k_p}{h} & -\frac{k_d}{h} & -k_d & -\frac{1}{h}
    \end{bmatrix},
    \, 
    B_w = \begin{bmatrix}
        1 & 0 \\
        0 & 0 \\
        0 & 0 \\
        \frac{k_d}{h} & \frac{1}{h}
    \end{bmatrix}
    \\
    B&= \begin{bmatrix}
        0 & 1 & 0 & 0 \\
        0 & 0 & 0 & 0 \\
        0 & 0 & 0 & 0 \\
        0 & \frac{k_d}{h} & 0 & \frac{1}{h}
    \end{bmatrix}, \, 
    B_e = \begin{bmatrix}
        0\\ 0\\ 0\\ \frac{1}{h}
    \end{bmatrix}, \,
    C_u = \begin{bmatrix}
     0 \\ 0 \\ 0 \\ 1
 \end{bmatrix}^\top.
\end{align*}

Next, the dynamics of the platoon at jumps, i.e. for every $t=t_k^+$, are given as follows: Since the considered vehicle dynamics are in continuous time, we have that $\mbf{x}(t_k^+)=\mbf{x}(t_k)$. 
On the other hand, the dynamics of the network-induced error $\mbf{e}$ at jumps are given by \eqref{eq:protocol-loss}. 
Therefore, by recalling that $\mbf{x}=(x_1,\dots,x_N)$, $\mbf{e}=(e_{u_1},\dots,e_{u_{N-1}})$ and $w=(v_0,u_0)$, we can write the following stochastic hybrid model for the platoon
\begin{align}\label{eq:HCS}
\mathcal{H}_1:\left\{
    \begin{aligned}
        \dot{\mbf{x}}(t) &= A_{11}\mbf{x}(t) + A_{12} \mbf{e}(t) + B_1w,\ t\in[t_{k-1},t_k], \\
    \dot{\mbf{e}}(t) &= A_{21}\mbf{x}(t) + A_{22}\mbf{e}(t)+B_2w,\ t\in[t_{k-1},t_k], \\
    \mbf{x}(t_k^+) &= \mbf{x}(t_k), \\
    \mbf{e}(t_k^+) &= \mbf{h}(k,\mbf{e}(t_k)),
    \end{aligned}
    \right.
\end{align}
with
\begin{align}\label{eq:matrices}
	A_{11} &= \smallmtx\begin{bmatrix}
	A &  &  &  & 0\\
	B & A & & &\\
	& \ddots & \ddots & & \\
	& & B & A &\\
	0 & & & B & A
	\end{bmatrix}\hspace{-1mm}, 
	A_{12} = \smallmtx\begin{bmatrix}
	0 &  &  & 0 \\
	B_e & 0 & & \\
	& \ddots & \ddots &  \\
	& & B_e & 0 \\
	0 & & & B_e 
	\end{bmatrix}\hspace{-1mm}, 
 B_1 = \begin{bmatrix}
		B_w\\ 0 \\ \vdots \\ 0 \\ 0
	\end{bmatrix}\hspace{-1mm},\nonumber\\
	%
 A_{21} &= \tilde{C}_u A_{11}, \ A_{22} = \tilde{C}_uA_{12},\  B_2 = \tilde{C}_uB_1, 
\end{align}
where $$\tilde{C}_u \coloneqq 
\begin{bmatrix}
	-C_u & 0 &  &  & 0\\
	0 & -C_u & & &\\
	& \ddots & \ddots & &\\
	0 & & 0 & -C_u &0\\
	\end{bmatrix}.$$

System $\mathcal{H}_1$ in \eqref{eq:HCS} captures the continuous dynamics given by the vehicles in the platoon (local controller and plant dynamics), and also the discrete stochastic dynamics (jumps) given by the network-induced effects,i.e., random packet loss, scheduling,  and stochastic transmission instants.

\subsection{String stability}
We highlight that the results in this section hold for a general class of non-linear cascaded systems given by
\begin{subequations}\label{eq:HCS-nonlinear}
	\begin{align}
		\dot{\mbf{x}}(t) &= f(\mbf{x}(t),\mbf{e}(t),w(t)), \ t\in[t_{k-1},t_k], \label{eq:x-sys} \\
		\dot{\mbf{e}}(t) &= g(\mbf{x}(t),\mbf{e}(t),w(t)), \ t\in[t_{k-1},t_k], \\
		\mbf{x}(t_k^+) &= \mbf{x}(t_k), \\
		\mbf{e}(t_k^+) &= \mbf{h}(k,\mbf{e}(t_k)),
	\end{align}    
\end{subequations}
where
\begin{align*}
	f(\mbf{x},\mbf{e},w) &\coloneqq \begin{bmatrix}
		f_1(x_1(t),w(t)) \\
		f_2(x_2(t),x_1(t),e_{u_1}(t)) \\
		\vdots \\
		f_N(x_N(t),x_{N-1}(t),e_{u_{N-1}}(t))
	\end{bmatrix},\\
	g(\mbf{x},\mbf{e},w) &\coloneqq \begin{bmatrix}
		g_1(x_1(t),w(t)) \\
		g_2(x_2(t),x_1(t),e_{u_1}(t)) \\
		\vdots \\
		g_{N-1}(x_{N-1}(t),x_{N-2}(t),e_{u_{N-2}}(t))
	\end{bmatrix},
\end{align*}
with $f_i$ and $g_i$ continuous non-linear functions. Note that \eqref{eq:HCS} is a particular case of \eqref{eq:HCS-nonlinear}.
Consequently, we first provide general string stability conditions that hold for  \eqref{eq:HCS-nonlinear}, and then in Section \ref{sec:explicit}, we show that more explicit  string stability conditions can be obtained for the linear case specified by \eqref{eq:HCS}.

The first step is to derive an input-output stability property in terms of $\EL_p$ gains with respect to relevant inputs and outputs of \eqref{eq:HCS-nonlinear}. We use the following property (cf. \cite{tabnes08-tac,maanes19}).

\begin{defn}\label{def:Lp-mean}
    For system \eqref{eq:HCS-nonlinear}, define $z\coloneqq (\mbf{x},\mbf{e})$ and consider any input and output functions $\mathscr{U}=F(z,w)$, $\mathscr{Y}=H(z,w)$, respectively. Let $p\in \N\cup\{+\infty\}$ and $\gamma\geq 0$ be given. We say that \eqref{eq:HCS-nonlinear} is \emph{$\EL_p$ stable in expectation}  from $\mathscr{U}$ to $\mathscr{Y}$ with gain $\tilde{\gamma}$ if there exists $\tilde{K}\geq 0$ such that any solution to \eqref{eq:HCS-nonlinear} with input $\mathscr{U}\in\EL_p^e $ satisfies
    \begin{align}\label{eq:Lp-exp}
        \espp{\norm{\mathscr{Y}}_{\EL_p[0,t]}} \leq \tilde{K}|z(0)| + \tilde{\gamma} \espp{\norm{\mathscr{U}}_{\EL_p[0,t]}},
    \end{align}
    for all $t\geq 0$.\fin 
\end{defn}

Definition \ref{def:Lp-mean} is the stochastic counterpart of the standard input-output or \emph{$\EL_p$ stability} property as defined, for example, in \cite{khalil02}. This represents an initial stage in the analysis of string stability since the constants $\tilde{K}$ and $\tilde{\gamma}$ in \eqref{eq:Lp-exp} are not (yet) required to be independent of the platoon length $N$.

A common approach in non-linear NCS analysis is to adopt a small-gain argument to provide conditions such that \eqref{eq:Lp-exp} holds. That is, system \eqref{eq:HCS-nonlinear} is decomposed into the interconnection between the $\mbf{x}$-- and $\mbf{e}$--subsystems, and the stability of this interconnection is ensured via small-gain arguments. This is formally stated in the following theorem.
\begin{theorem}\label{theo:Lp-overall}
	Consider system \eqref{eq:HCS-nonlinear} and suppose the following conditions hold.
	\begin{enumerate}[$(i)$]
		\item Assumption \ref{assu:as-UGES} holds with a Lyapunov function $W$ that is locally Lipschitz in $\mbf{e}$, uniformly in $k$.
		\item There exists $L\in\R_{\geq 0}$ such that
		\begin{align}\label{eq:dWdt}
		\left\langle \partial W/\partial \mbf{e},g(\mbf{x},\mbf{e},w)\right\rangle \leq LW(k,\mbf{e}) + |\tilde{y}(\mbf{x},w)|
		\end{align}
		holds for almost all $\mbf{e}\in\R^{n_e}$, all $(\mbf{x},w)\in\R^{n_x}\times\R^{n_w}$, $t\in(t_k,t_{k+1})$, and $k\in\N_0$, where $\tilde{y}\coloneqq G(\mbf{x})+ Ew$, for some $G:\R^{n_x}\rightarrow \R^{n_e}$ and $E\in\R^{n_e\times n_w}$.
		\item The $\mbf{x}$--subsystem  \eqref{eq:x-sys} is $\mathcal{L}_p$ stable in expectation from $(W(\mbf{e}),w)$ to $G(\mbf{x})$ with gain $\gamma_x$ according to Definition \ref{def:Lp-mean}, for some $p\in\N\cup\{+\infty\}$ and $K_x\geq 0$.
	\end{enumerate}
	If the transmission rate $\w$ satisfies
	\begin{align}\label{eq:omega-closed}
	\w > (\gamma_x + L)/(1 - \bar{\kappa}),
	\end{align}
	then, system \eqref{eq:HCS-nonlinear} is $\EL_p$ stable in expectation from $w$ to $(G(\mbf{x}),W(\mbf{e}))$ as per Definition \ref{def:Lp-mean}, with
 \begin{align*}
     \tilde{\gamma} &=\frac{\gamma_x + \gamma_e|E| +  (1+|E|)\gamma_x\gamma_e}{1 - \gamma_x \gamma_e}, \\
     \tilde{K} &= \max\{K_x(1+\gamma_x),K_e(1+\gamma_e)\}\frac{\sqrt{\max\{1+\epsilon,1+1/\epsilon\}}}{1-\gamma_x\gamma_e},
 \end{align*}
  where $\epsilon>0$, $K_e = \tfrac{\overline{a}_W(\w - L)}{\max\{1,L\}(\w(1-\bar{\kappa})-L)}$, and $\gamma_e = \tfrac{1}{\w (1 - \bar{\kappa}) -L}$.	
\end{theorem}
\textbf{Proof:} See Appendix \ref{sec:proof-small-gain}.\qed 

Theorem \ref{theo:Lp-overall} provides a lower bound for the transmission rate $\w$ that ensures input-output stability of system \eqref{eq:HCS-nonlinear}. This property takes us one step closer to string stability, but further analysis is needed. Similar to \cite{monteil2018,ploeg2013lp}, if the  input-output property from Theorem \ref{theo:Lp-overall} holds for any platoon length $N$ (i.e., $\tilde{K}$ and $\tilde{\gamma}$ do not depend on $N$ or are upper bounded by constants independent of $N$), then we are able to ensure string stability as per Definition \ref{def:Lp-MSS}.

All the conditions in Theorem \ref{theo:Lp-overall} can be guaranteed to hold. Condition $(i)$ relates to the implemented scheduling protocols. Many protocols available in the literature satisfy this property, see e.g. \cite{nestee04}; and we have provided two examples earlier (RR and SD protocols). Condition $(ii)$ assumes an exponential growth on the $\mbf{e}$--subsystem and it is satisfied when $W$ is globally Lipschitz in $\mbf{e}$ uniformly in $k$, and $g$ is globally Lipschitz for instance, see \cite{nestee04}.
Condition $(iii)$ ensures the controller has been designed to achieve $\EL_p$ stability of the network-free $\mbf{x}$--subsystem. This is typically ensured in the first step of emulation.
Since, for any stabilisable and detectable LTI system, this property holds true.

\begin{remark}\label{rem:SSS}
 Note that $\tilde{K}$ and $\tilde{\gamma}$ in Theorem \ref{theo:Lp-overall} depend on the parameters $\{K_x,\gamma_x,\overline{a}_W,\w,L,\bar{\kappa},|E|\}$. It is important to recall that our objective is to attain string stability, with these parameters independent of $N$. These parameters correspond to various aspects of the platoon, and we provide the following remarks concerning each of them.

\noindent\textbf{Network-free string stability:} The parameters $\{\gamma_x, K_x\}$ are associated with the string stability of the network-free platoon. Typically, these parameters are independent of $N$ (or upper bounded by constants independent of $N$) when the controller is appropriately designed in the absence of a network.

 \noindent \textbf{Transmission rate:} This parameter $\w$ does not depend on $N$ by definition, see Assumption \ref{assu:transmission-instants}. 

 \noindent\textbf{Platoon dynamics and topology:} The parameters $\{L, |E|\}$ are associated with the configuration and dynamics of the platoon, and, as such, they depend on $N$ beforehand. Various choices for the platoon setting will result in different values of $L$ and $|E|$. However, in Section \ref{sec:explicit}, we show that for the platoon configuration outlined in Sections \ref{sec:setup} and \ref{sec:network}, these parameters are indeed independent of $N$. Similar analysis holds for other possible configurations.

 \noindent\textbf{Scheduling protocols:} The parameters $\{\overline{a}_W, \bar{\kappa}\}$ are linked to the scheduling protocol employed. In protocols like RR, these parameters often vary with $N$, while in  SD protocol, such dependence is absent. Therefore, the protocol choice is crucial for achieving string stability. This observation aligns with the idea that TDMA-based protocols (such as RR) may not be the best choice for platooning as they  introduce delays and limitations due to divided time slots, hindering real-time communication and coordination required for maintaining consistent spacing and velocity among vehicles \cite{aslam2018flexible}. Nevertheless, it is important to clarify that this paper does not aim to conduct a comprehensive analysis of scheduling protocols and their impact on platooning. Rather, our focus is to shed light on the influence of these protocols on string stability by examining specific parameters.\fin 
\end{remark}

Based on the guidelines provided in Remark \ref{rem:SSS}, we can now present the following corollary on string stability for the non-linear interconnected platoon described by \eqref{eq:HCS-nonlinear}.

\begin{coro}\label{coro:SS-nolinear}
	Consider system \eqref{eq:HCS-nonlinear} under conditions $(i)$--$(iii)$ in Theorem \ref{theo:Lp-overall}. If the following holds,
 \begin{enumerate}[$(i)$]
     \item The time-headway $h$ is designed such that there exist $ \overline{\gamma}_x,\overline{K}_x\geq 0$ such that $\gamma_x\leq \overline{\gamma}_x$ and $K_x\leq \overline{K}_x$ for any $N\in\N$, where $\gamma_x,K_x$ are as per Theorem \ref{theo:Lp-overall}$(iii)$.
     \item The transmission rate satisfies $\w>(\overline{\gamma}_x+L)/p$.
     \item The SD protocol in Example \ref{ex:SD} is used to schedule V2V communications.
     \item The platoon configuration is such that $L$ and $|E|$ in \eqref{eq:dWdt} are independent of $N$.
     \item System \eqref{eq:HCS-nonlinear} is $\EL_p$ to $\EL_p$ detectable in expectation from $(G(\mbf{x}),W(\mbf{e}))$ to $(\mbf{x},\mbf{e})$. That is, there exists $K,\gamma\geq 0$ such that any solution to \eqref{eq:HCS-nonlinear} with input $w\in\EL_p$ verifies $
        \espp{\norm{\mbf{(x,e)}}_{\EL_p[0,t]}} \leq K|(\mbf{x}(0),\mbf{e}(0))| + \gamma\norm{(G(\mbf{x}),W(\mbf{e}))}_{\EL_p[0,t]} + \gamma \norm{w}_{\EL_p[0,t]}$,
    for all $t\geq 0$ and some $p\in\N\cup\{+\infty\}$.
 \end{enumerate}
 Then, system \eqref{eq:HCS-nonlinear} is $\EL_p$ string stable in expectation.\fin 
\end{coro}

Conditions $(i)$--$(iv)$ follow from the guidelines outlined in Remark \ref{rem:SSS}. Condition $(v)$ is a detectability assumption on the interconnected system \eqref{eq:HCS-nonlinear}, see e.g. \cite{nestee04}. It is worth mentioning that Corollary \ref{coro:SS-nolinear} applies to the general cascaded system \eqref{eq:HCS-nonlinear}, however,  it does not provide explicit conditions for $\EL_p$ string stability in expectation in terms of specific vehicle and network parameters. In the case of the linear platoon \eqref{eq:HCS}, we show in the following section that  closed-form conditions can be obtained that explicitly relate key vehicle and network parameters, ensuring string stability in expectation.


\section{Explicit string stability conditions}\label{sec:explicit}
Our attention now shifts to the linear platoon $\mathcal{H}_1$ described by \eqref{eq:HCS}, allowing us to establish more explicit conditions that guarantee $\EL_p$ string stability in expectation for $p=2$. 
Formally, sufficient conditions for $\EL_2$ string stability in expectation of $\mathcal{H}_1$ are stated in the following theorem.
\begin{theorem}\label{theo:SS-explicit}
    Consider the wireless platoon $\mathcal{H}_1$ in \eqref{eq:HCS} under the SD protocol in Example \ref{ex:SD}, and let $P(s)\coloneqq A_{21}(sI-A_{11})^{-1}[A_{12}\ B_1]$ be the transfer function of the $\mbf{x}$--subsystem in \eqref{eq:HCS}, with $(A_{11},A_{21})$ detectable. If the following holds
    \begin{enumerate}[$(i)$]
        \item The time-headway $h$ is such that there exists $\overline{\gamma}_x,\overline{K}_x\geq 0$ 
satisfying $\norm{P(j\omega)}_{\mathcal{H}_\infty}\leq \overline{\gamma}_x$ and $|A_{21}|\leq \overline{K}_x$  for any platoon length $N\in\N$.
        \item The transmission rate satisfies
    \begin{align}\label{eq:condition-sss}
        \w > \frac{1}{\alpha}\left(\overline{\gamma}_x + \frac{1}{h}\right).
    \end{align}
    \end{enumerate}
    Then, $\mathcal{H}_1$ is $\EL_2$ string stable in expectation. 
\end{theorem}
\textbf{Proof:} See Appendix \ref{sec:proof-SS-explicit}.\qed

Condition $(i)$ is related to the network-free design of $h$, meaning that the platoon should exhibit string stability in the absence of network imperfections. Furthermore, condition $(ii)$ establishes a direct relationship between the parameters of the platoon: the probability of successful transmission $\alpha$, the transmission rate $\w$, and the time-headway $h$. 

From the condition \eqref{eq:condition-sss}, it can be noticed that in order to achieve $\EL_2$ string stability in expectation when employing lower-quality channels, the time-headway $h$ should be increased. The observation made in \cite{vargas2018} through simulations, albeit in a slightly different scenario, only hinted at this discovery. However, Theorem \ref{theo:SS-explicit} offers a theoretical foundation to support this finding.
Furthermore, we examine the influence of the transmission rate $\w$ on stochastic string stability, an aspect that was not investigated in \cite{vargas2018}, even in simulation studies. As observed from \eqref{eq:condition-sss}, in the case of a low-quality channel, achieving string stability relies not only on increasing the time-headway $h$ but also on having more frequent transmissions. Additionally, we observe that maintaining the same value for $h$ as in the network-free case can still lead to achieving string stability under packet losses by increasing the transmission rate, provided the network has enough capacity.

In the following, we illustrate the utility of the proposed tools in a different platooning configuration often seen in practice (cf. Remark \ref{rem:local-sensors}), where we transmit not only $u_{i-1}$ over the network, as shown in Figure \ref{fig:block_diagram}, but also the predecessor's position and velocity $(s_{i-1}, v_{i-1})$. This particular case has also been explored in the literature, but in the context of event-triggered control \cite{li2019string}.
The additional information becomes relevant when local sensors are allocated for other purposes such as obstacle avoidance or redundancy, as discussed in Remark \ref{rem:local-sensors}. 
In this new setting, the spacing error between $\mathcal{V}_i$ and $\mathcal{V}_{i-1}$ is given by $\widehat{\xi}_i \coloneqq [\hat{s}_{i-1}-s_i-L_i]-[r_i+hv_i]=\xi_i + e_{s_{i-1}}$, where
$\hat{s}_{i-1}$ denotes the transmitted version of the signal $s_{i-1}$,
$\xi_i$ is the network-free spacing error and $e_{s_{i-1}}\coloneqq \hat{s}_{i-1}-s_{i-1}$ is the network-induced error associated with the transmission of $s_{i-1}$. Similarly, since $v_{i-1}$ is also transmitted, we define $\widehat{\nu}_i\coloneqq \hat{v}_{i-1} - v_i - ha_i=\dot{\xi}_i + e_{v_{i-1}}$, where $\dot{\xi}_i$ is the network-free derivative of the spacing error, and $e_{v_{i-1}}$ denotes the network-induced error associated with the transmission of $v_{i-1}$. Therefore, the 
control law in this new setting takes the form, for all $t\geq 0$ and $i\in\mathcal{N}$,
\begin{multline}\label{eq:controller-new}
    q_i(t) = \underbrace{k_p \xi_i(t) + k_d \dot{\xi}_i(t) + u_{i-1}(t)}_{\mathrm{network-free\ control\ law}}\\ + \underbrace{k_pe_{s_{i-1}}(t) + k_d e_{v_{i-1}}(t) + e_{u_{i-1}}(t)}_{\mathrm{network-induced\ errors}} .
\end{multline}
Let $\mbf{e}_{i-1}\coloneqq  (e_{s_{i-1}},e_{v_{i-1}},e_{u_{i-1}})$ and $\tilde{\mbf{e}}\coloneqq (\mbf{e}_1,\dots,\mbf{e}_{N-1})\in\R^{n_{\tilde{e}}}$, where $n_{\tilde{e}}=3(N-1)$. Then, proceeding similarly to Section \ref{sec:hybrid-model}, we can describe this scenario using the following hybrid model
\begin{align}\label{eq:HCS-2}
\mathcal{H}_2:\left\{
    \begin{aligned}
        \dot{\mbf{x}}(t) &= A_{11}\mbf{x}(t) + \tilde{A}_{12} \tilde{\mbf{e}}(t) + B_1w,\ t\in[t_{k-1},t_k], \\
    \dot{\tilde{\mbf{e}}}(t) &= \tilde{A}_{21}\mbf{x}(t) + \tilde{A}_{22}\tilde{\mbf{e}}(t)+\tilde{B}_2w,\ t\in[t_{k-1},t_k], \\
    \mbf{x}(t_k^+) &= \mbf{x}(t_k), \\
    \tilde{\mbf{e}}(t_k^+) &= \mbf{h}(k,\tilde{\mbf{e}}(t_k)),
    \end{aligned}
    \right.
\end{align}
with $(A_{11},B_1)$ as in \eqref{eq:matrices} and $(\tilde{A}_{12},\tilde{A}_{21},\tilde{A}_{22},\tilde{B}_2)$ given by
\begin{align}\label{eq:matrices-new}
	\tilde{A}_{12} &= \smallmtx\begin{bmatrix}
	0 &  &  & 0 \\
	B_eK_e & 0 & & \\
	& \ddots & \ddots &  \\
	& & B_eK_e & 0 \\
	0 & & & B_e K_e
	\end{bmatrix}, \\
 \tilde{A}_{21} &= \tilde{C}A_{11}+ \bar{C}, \quad
 \tilde{A}_{22} = \tilde{C}A_{12},\ \tilde{B}_2 = \tilde{C}B_1,\nonumber
 \end{align}
 where
 \begin{align*}
     K_e &\coloneqq \begin{bmatrix} k_p & k_d & 1\end{bmatrix}^\top, \
     \tilde{C}\coloneqq \begin{bmatrix}
            C_1^\top & \cdots & C_{N-1}^\top
        \end{bmatrix}^\top, \\
        C_1 &\coloneqq \begin{bmatrix}
            0 & -C_s & \cdots & 0 \\
            -C_v & 0 & \cdots & 0 \\
            -C_u & 0 & \cdots & 0
        \end{bmatrix}, C_s\coloneqq \begin{bmatrix}
            1 & 0 & 0 & 0
        \end{bmatrix},\\
 C_{N-1} &\coloneqq \begin{bmatrix}
     0 & \cdots & 0 & -C_s \\
     0 & \cdots & -C_v & 0 \\
     0 & \cdots & -C_u & 0
 \end{bmatrix}, C_v\coloneqq \begin{bmatrix}
            0 & 1 & 0 & 0
        \end{bmatrix}, \\
\bar{C}_\iota &\coloneqq \begin{bmatrix}
    0 & -C_v-hC_a & \cdots & 0 \\
    0 & 0 & \cdots & 0\\
        0 & 0 & \cdots & 0
\end{bmatrix}, C_a \coloneqq \begin{bmatrix}
            0 & 0 & 1 & 0
        \end{bmatrix},\\
        \bar{C} &\coloneqq \begin{bmatrix}
            \bar{C}_\iota^\top & \cdots & \bar{C}_\iota^\top
        \end{bmatrix}^\top ,
        C_u \coloneqq \begin{bmatrix}
            0 & 0 & 0 & 1
        \end{bmatrix}.
 \end{align*}

The theorem below presents the corresponding string stability conditions for this case. 
\begin{theorem}\label{theo:SS-explicit-new}
    Consider the wireless platoon $\mathcal{H}_2$ in \eqref{eq:HCS-2} under the SD protocol in Example \ref{ex:SD}, and let $Q(s)\coloneqq \tilde{A}_{21}(sI-A_{11})^{-1}[\tilde{A}_{12}\ B_1]$ be the corresponding transfer function of the $\mbf{x}$--subsystem, with $(A_{11},\tilde{A}_{21})$ detectable. If the following holds
    \begin{enumerate}[$(i)$]
        \item The time-headway $h$ is such that there exists $\overline{\vartheta}_x,\overline{M}_x\geq 0$ 
satisfying $\norm{Q(j\omega)}_{\mathcal{H}_\infty}\leq \overline{\vartheta}_x$ and $|\tilde{A}_{21}|\leq \overline{M}_x$  for any platoon length $N\in\N$.
        \item The transmission rate satisfies
    \begin{align}\label{eq:condition-sss-new}
        \w > \frac{1}{\alpha}\left(\overline{\vartheta}_x + \frac{1}{h}\sqrt{1 + k_d^2 + k_p^2}\right).
    \end{align}
    \end{enumerate}
    Then, $\mathcal{H}_2$ is $\EL_2$ string stable in expectation.
\end{theorem}
\textbf{Proof:} See Appendix \ref{sec:proof-SS-explicit-2}.\qed

It can be seen from \eqref{eq:condition-sss-new} that, in this new scenario, the  controller parameters $k_d$ and $k_p$ also play a role in the success probability bound for achieving string stability. Specifically, \eqref{eq:condition-sss-new} might result in a reduced stability region when compared to \eqref{eq:condition-sss}.  This effect could have been anticipated since we are transmitting more data signals over the network, making them susceptible to packet losses.

\section{Numerical Example}\label{sec:examples}
Consider the wireless platoon described by $\mathcal{H}_1$ under the SD protocol, with controller gains $k_p=0.2$ and $k_d=0.7$, and drive-line constant $\tau=0.1$. These parameter values achieve individual vehicle stability as shown in \cite{ploeg2013lp}. The objective of this example is to illustrate how the interplay among key platoon parameters, such as $h$, $\alpha$, and $\w$, directly impact string stability.

\subsection{Theoretical bounds for string stability}
\begin{figure}[htb]
	\begin{center}
		\includegraphics[scale=0.55]{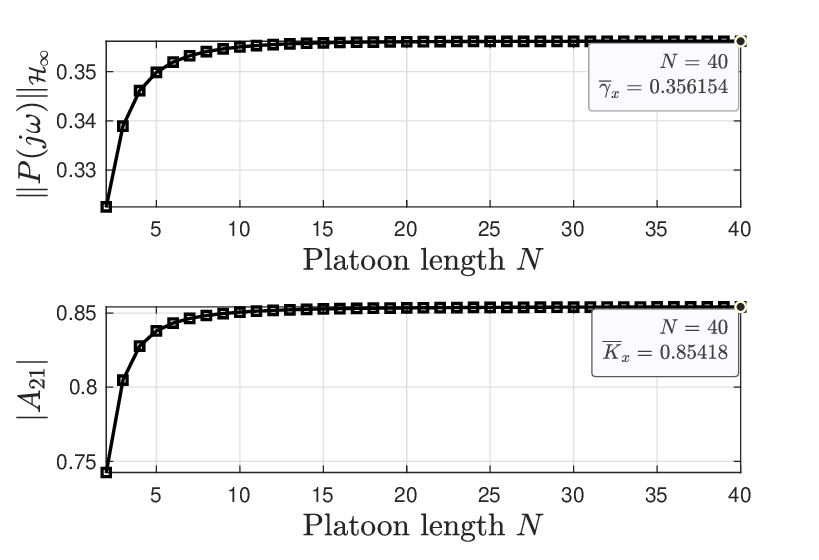}       
		\caption{Upper bounds $\overline{\gamma}_x$ and $\overline{K}_x$ as per Theorem \ref{theo:SS-explicit}$(i)$ for $h=5$.}  
		\label{fig:upper-bounds}                      \end{center}
  \end{figure}

\begin{figure}[htb]
	\begin{center}
		\includegraphics[scale=0.55]{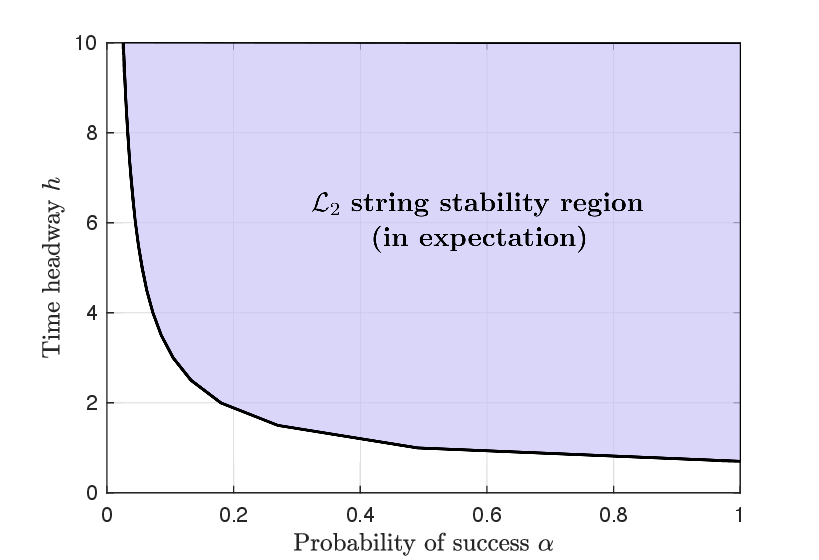}       
		\caption{Region for $\EL_2$ string stability in expectation as per Theorem \ref{theo:SS-explicit}, with an average transmission interval of $1/\w=0.1[s]$.}  
		\label{fig:region}                      \end{center}
  \end{figure}

We first illustrate how the explicit conditions in Theorem \ref{theo:SS-explicit} lead to a $(h,\alpha)$ region that ensures string stability in expectation for $\mathcal{H}_1$. Let $\w=10$, which means we have an average inter-transmission period of $0.1[s]$. To compute $\overline{\gamma}_x$, we
emphasise that each value of $h$ will lead to a different value for $\overline{\gamma}_x$ in Theorem \ref{theo:SS-explicit}, as the $\mbf{x}$--subsystem transfer function $P(s)$ depends on $h$. Figure \ref{fig:upper-bounds} illustrates this fact, where we can see that $\norm{P(j\omega)}_{\mathcal{H}_{\infty}}$ and $|A_{21}|$ are indeed upper bounded by constants $\overline{\gamma}_x=0.356$ and $\overline{K}_x=0.854$ which are independent of the platoon length $N$, for $h=5$. We can thus repeat this process to find the corresponding $\overline{\gamma}_x$ for each $h$ and use \eqref{eq:condition-sss} to plot the string stability region in Figure \ref{fig:region}. We can see that, as expected, lower quality channels (i.e. small $\alpha$) require a larger headway $h$ in order to ensure string stability. Note that our theoretical bounds are sufficient conditions. Thus, in practice, a smaller $h$ may still lead to string stability. In the case of higher-quality channels (large $\alpha$), a small value of $h$ would suffice to achieve string stability. Additionally, different values of transmission rate $\w$ may also help achieving string stability, as illustrated in the following section.

\begin{figure}[htb]
	\begin{center}
		\includegraphics[scale=0.55]{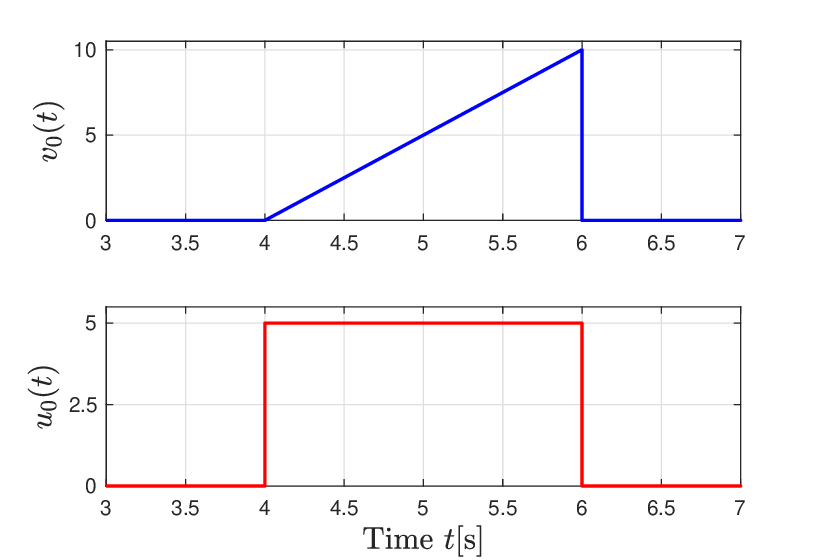}       
		\caption{External platoon input $w(t)=(v_0(t),u_0(t))$.}  
		\label{fig:input}                      \end{center}
  \end{figure}

\subsection{String stability simulations}
\begin{figure*}[!htb]\centering
\begin{tabular}{cc}
   \begin{minipage}{0.45\textwidth}
\includegraphics[width=0.8\linewidth]{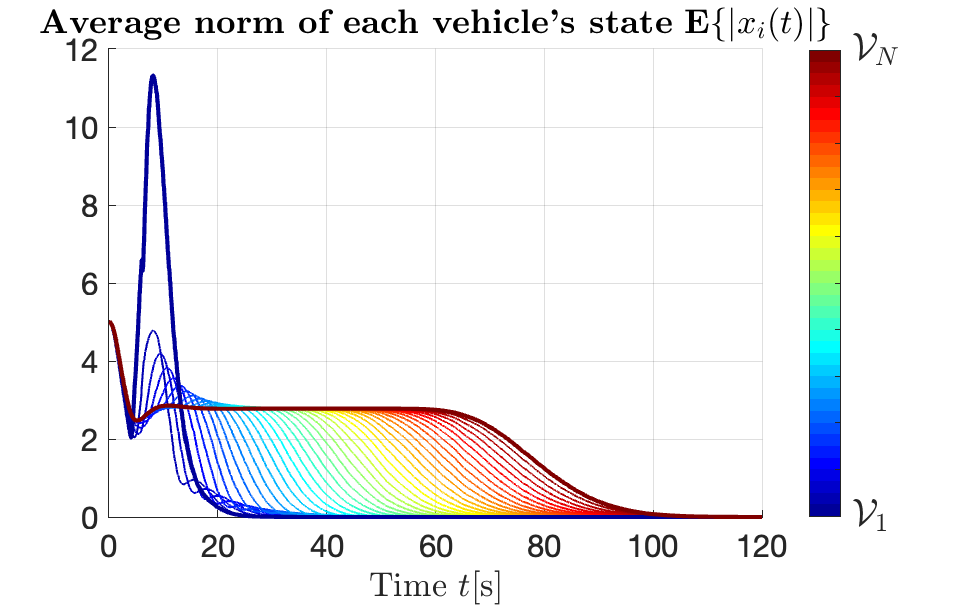}
\end{minipage}
& 
   \begin{minipage}{0.45\textwidth}
\includegraphics[width=0.8\linewidth]{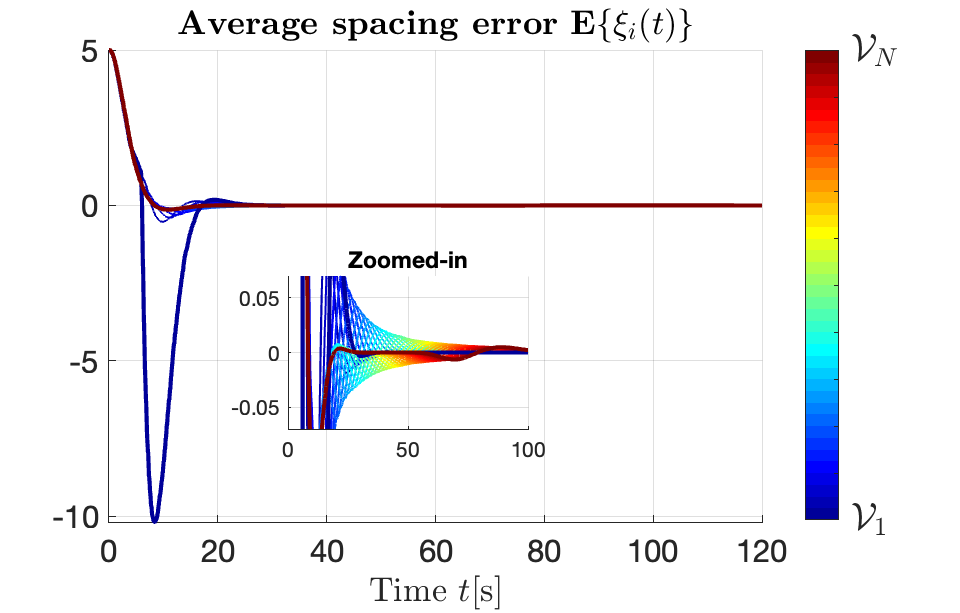}
   \end{minipage} \\
      \begin{minipage}{0.45\textwidth}
\includegraphics[width=0.8\linewidth]{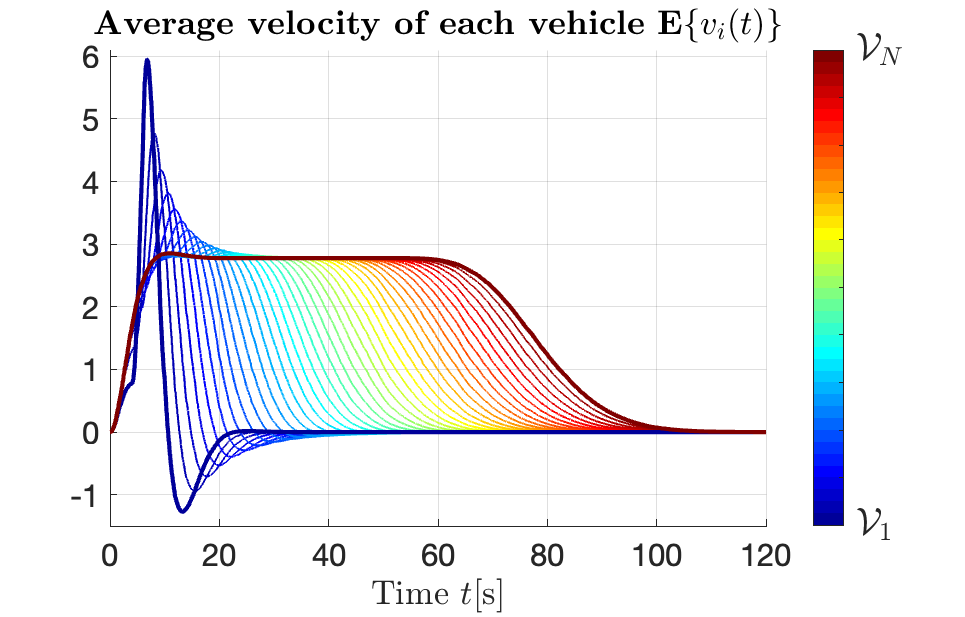}
\end{minipage}
& 
   \begin{minipage}{0.45\textwidth}
\includegraphics[width=0.8\linewidth]{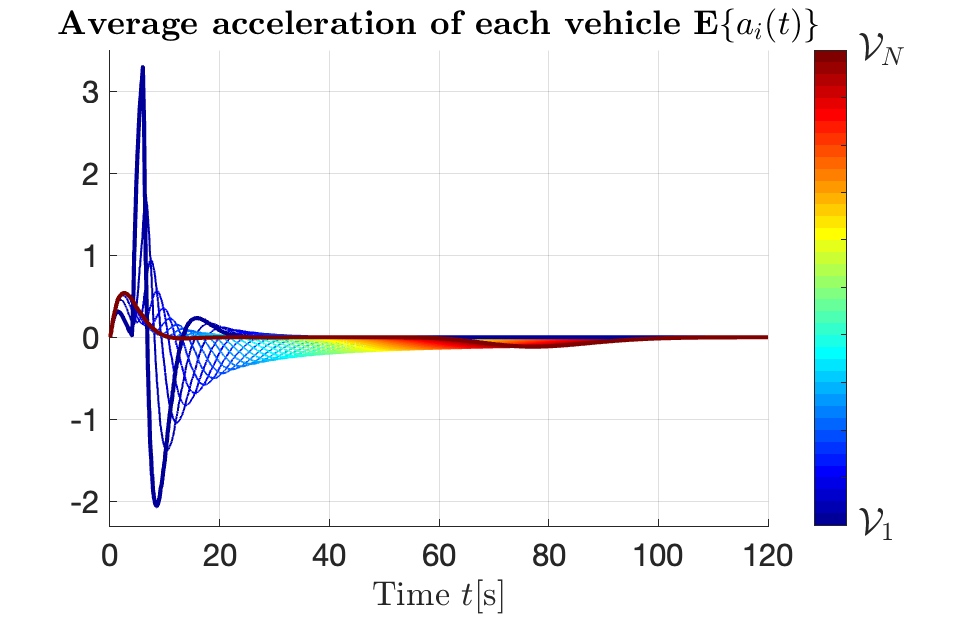}
   \end{minipage}
\end{tabular}
\caption{Expected values for relevant signals in the platoon of $N=40$ vehicles over 300 realizations with $h=1.8$, $1/\w=0.1$ and $\alpha=0.5$.}
\label{fig:signals-string-stable-via-MATI}
\end{figure*}

Here, we use the Hybrid Equations Toolbox (HyEQ) \cite{sanfelice2013toolbox} to simulate the stochastic hybrid system $\mathcal{H}_1$ under different scenarios of interest, to illustrate the interaction of various platoon parameters and how they contribute to achieving string stability (in expectation). This will serve to validate our theoretical findings. The velocity and acceleration profile that determines the external $\EL_2$ input $w=(v_0,u_0)$ are given in Figure \ref{fig:input}. 
The initial condition for every vehicle is taken to be $x_i(0)=(5,0,0,0)$.
We first show that, for a transmission rate of $\w=10$, the pair $(h,\alpha)=(1.8,0.5)$ leads to string stability for a platoon of $N=40$ vehicles, as ensured by the theoretical region in Figure \ref{fig:region}. Indeed, Figure \ref{fig:signals-string-stable-via-MATI} represents this scenario where the expected values of four significant signals, including norms of vehicle state, spacing errors, velocities, and accelerations are shown. We can see that the platoon is $\EL_2$ string stable in expectation as per Definition \ref{def:Lp-MSS}.

\begin{figure*}[!htb]\centering
\begin{tabular}{cc}
   \begin{minipage}{0.45\textwidth}
\includegraphics[width=0.8\linewidth]{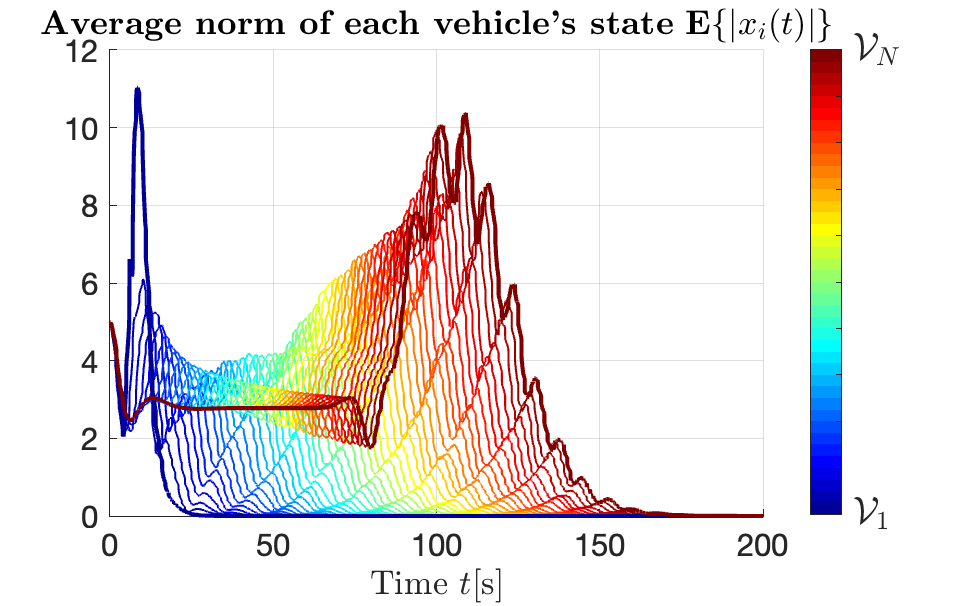}
\end{minipage}
& 
   \begin{minipage}{0.45\textwidth}
\includegraphics[width=0.8\linewidth]{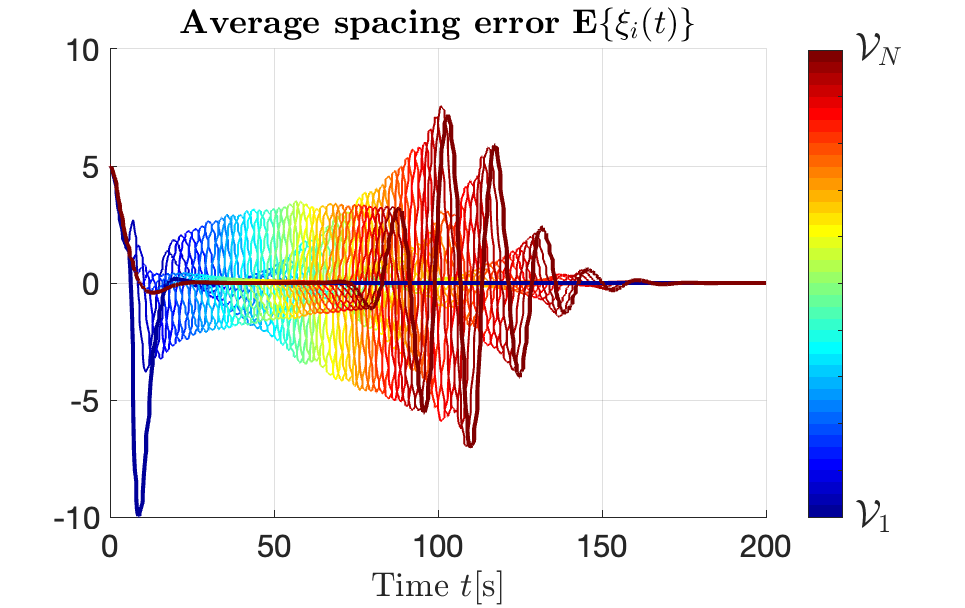}
   \end{minipage} \\
      \begin{minipage}{0.45\textwidth}
\includegraphics[width=0.8\linewidth]{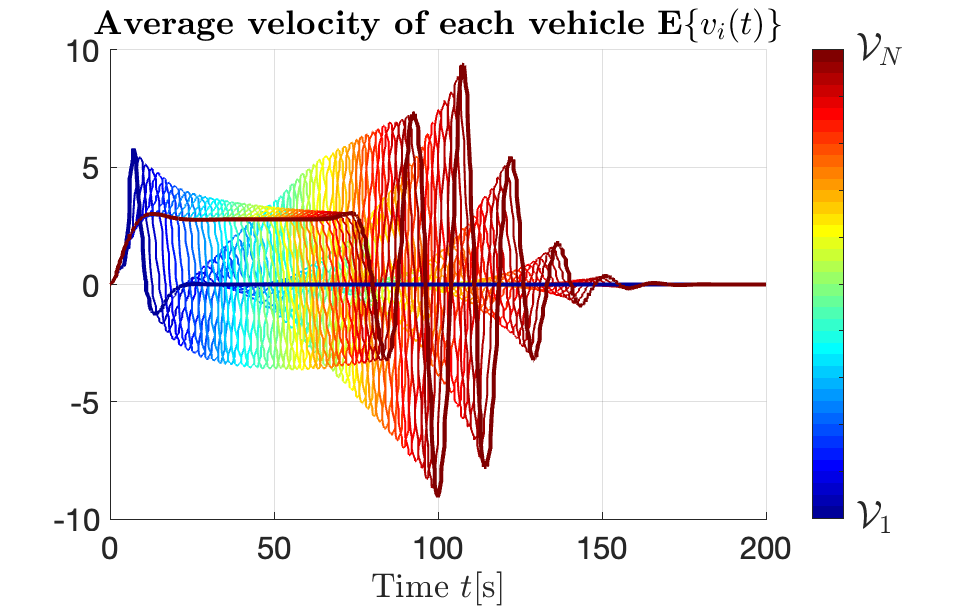}
\end{minipage}
& 
   \begin{minipage}{0.45\textwidth}
\includegraphics[width=0.8\linewidth]{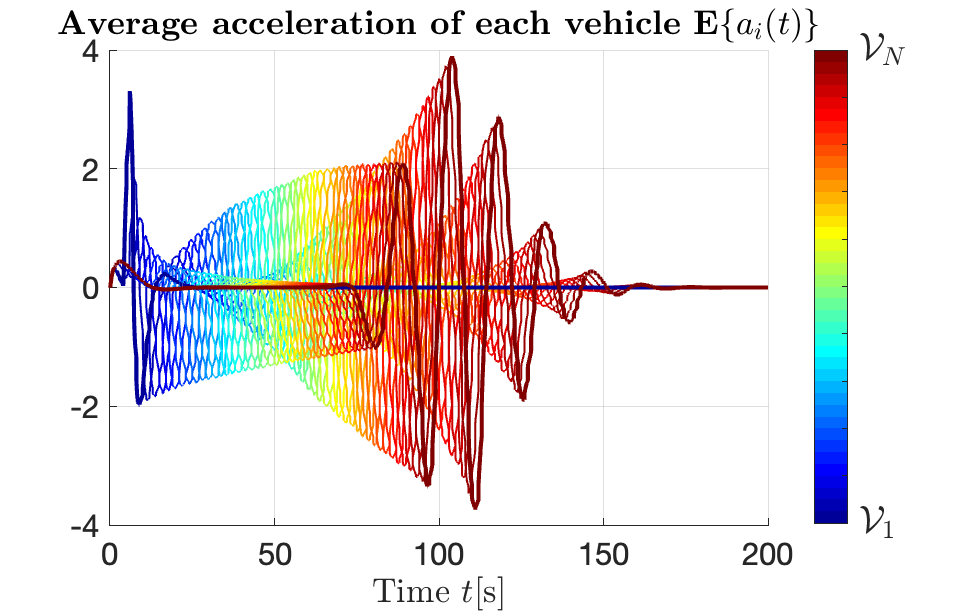}
   \end{minipage}
\end{tabular}
\caption{Expected values for relevant signals in the platoon of $N=40$ vehicles over 1000 realizations with $h=1.8$, $1/\w=1$ and $\alpha=0.5$.}
\label{fig:signals-string-unstable}
\end{figure*}

The next scenario shown in Figure \ref{fig:signals-string-unstable} illustrates string instability for the same pair $(h,\alpha)=(1.8,0.5)$, but with a slower transmission rate of $\w=1$, i.e. transmissions every $1[s]$ in average (compared to every $0.1[s]$ in the previous case). We can see that the frequency of transmissions has also an impact in achieving string stability as our theoretical results already anticipated.

\begin{figure*}[!htb]\centering
\begin{tabular}{cc}
   \begin{minipage}{0.45\textwidth}
\includegraphics[width=0.8\linewidth]{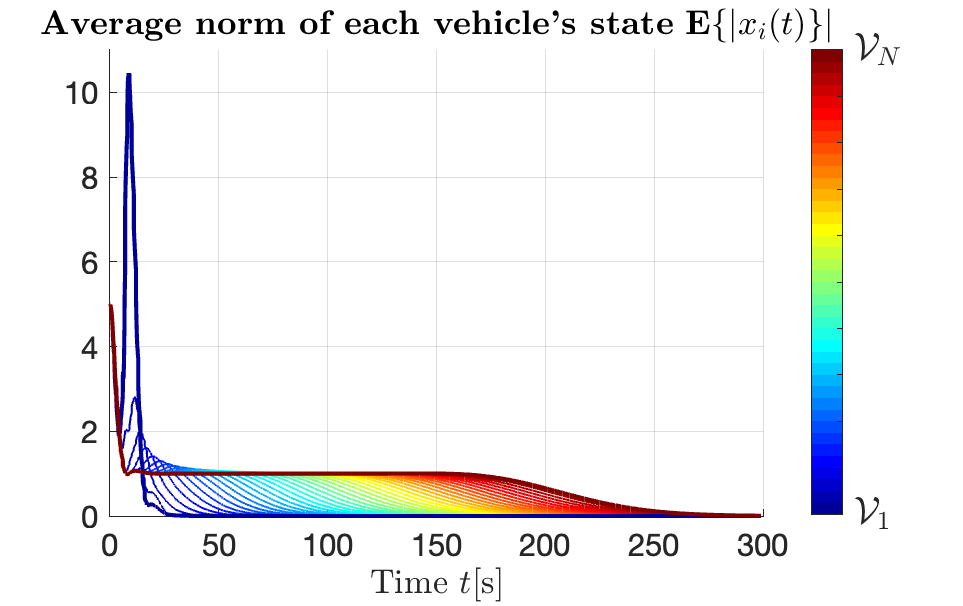}
\end{minipage}
& 
   \begin{minipage}{0.45\textwidth}
\includegraphics[width=0.8\linewidth]{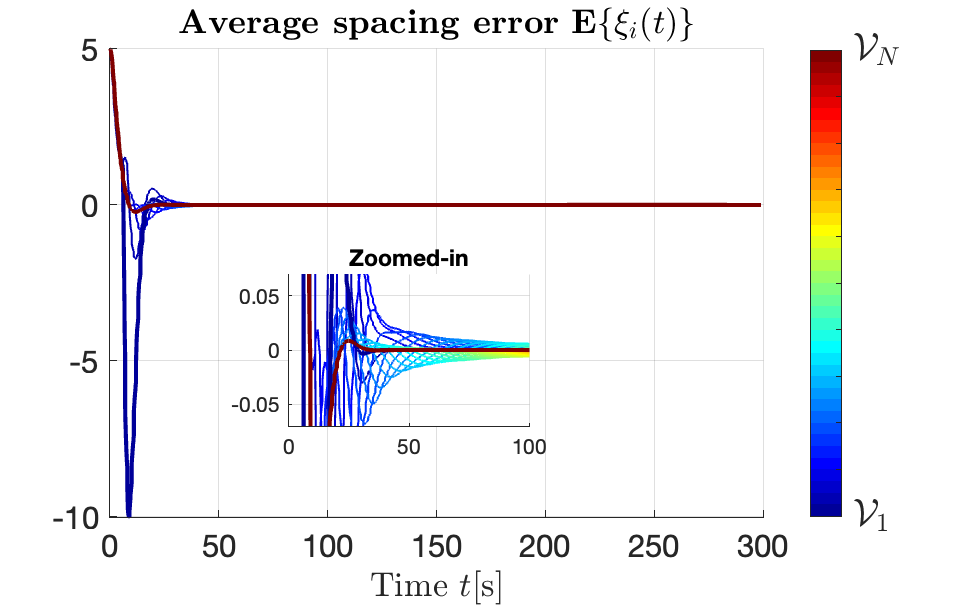}
   \end{minipage} \\
      \begin{minipage}{0.45\textwidth}
\includegraphics[width=0.8\linewidth]{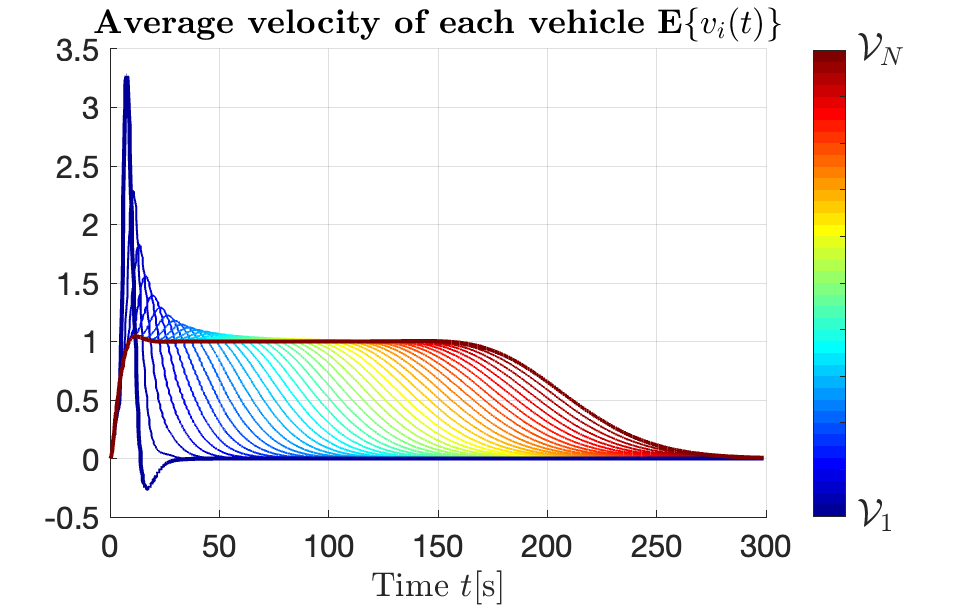}
\end{minipage}
& 
   \begin{minipage}{0.45\textwidth}
\includegraphics[width=0.8\linewidth]{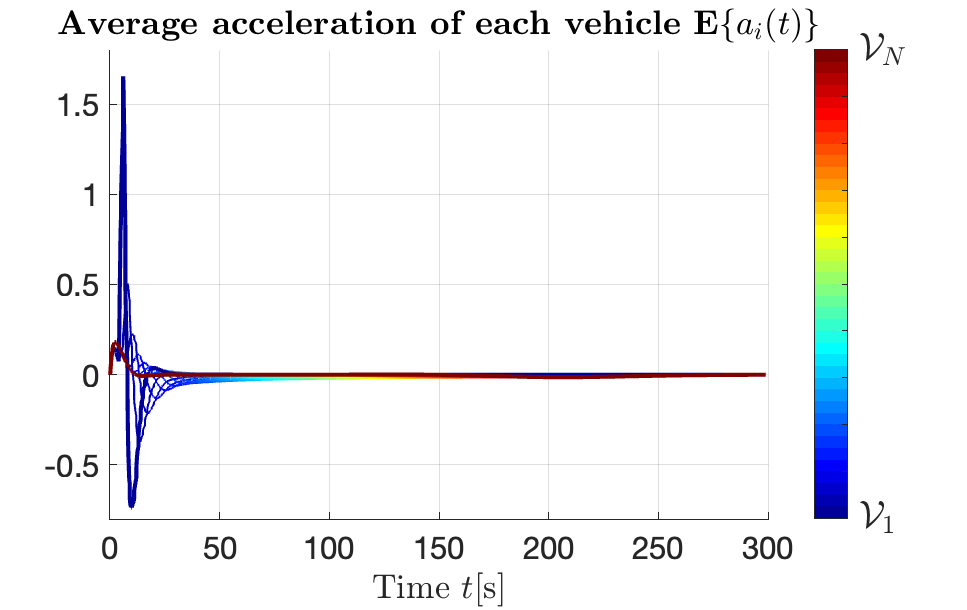}
   \end{minipage}
\end{tabular}
\caption{Expected values for relevant signals in the platoon of $N=40$ vehicles over 300 realizations with $h=5$, $1/\w=1$ and $\alpha=0.5$.}
\label{fig:signals-string-stable-via-h}
\end{figure*}

Finally, we illustrate that, even when the communication channel only allows for slower transmission rates, e.g. $\w=1$, we can still achieve string stability by adjusting the headway $h$. For instance, in Figure \ref{fig:signals-string-stable-via-h}, we can see the platoon is $\EL_2$ string stable in expectation for $(h,\alpha)=(5,0.5)$.

\section{Conclusion}
In this paper, we provided a framework based on hybrid systems to analyse string stability of predecessor-following platoons under packet losses. Importantly, we  obtained theoretical bounds that show how the interplay among key parameters of the platoon, such as time headway, channel probability, and transmission rate, impact string stability in expectation. These results were also illustrated via numerical examples. 
We believe that these findings have important practical implications for improving platooning systems under communication uncertainties.
Future work includes exploring 
different probabilities for each link in the platoon, diverse information flow topologies, and heterogeneous platoons.


\appendix 
\subsection{Proof of Theorem \ref{theo:Lp-overall}}\label{sec:proof-small-gain}
We consider system \eqref{eq:HCS-nonlinear} as the interconnection of two subsystems, namely the $\mbf{x}$--subsystem and the $\mbf{e}$--subsystem.
We first show an input-output stability property for the $\mbf{e}$--subsystem, and since we assume a input-output stability property for the $\mbf{x}$--subsystem, we use small-gain arguments to conclude the proof.

To show the input-output stability property for the $\mbf{e}$--subsystem we use Proposition 9.5 in \cite{tabnes08-tac}, which applies to general systems of the form \eqref{eq:HCS-nonlinear}.
Note that \eqref{eq:omega-closed} implies that $\w > L/(1-\bar{\kappa})$, and since conditions $(i)$ and $(ii)$ of the theorem hold, we can conclude from Proposition 9.5 in \cite{tabnes08-tac} that
\begin{multline}\label{eq:sg-e}
	\espp{\norm{W(\mbf{e})}_{\EL_p[t_0,t]}} \leq K_e|\mbf{e}(0)|\\ + \gamma_e \espp{ \norm{\tilde{y}(\mbf{x},w)}_{\EL_p[t_0,t]} },
\end{multline}
with $K_e$ and $\gamma_e$ as defined in the theorem statement.

Now, from Condition $(iii)$, we have that there exist $K_x\geq 0$ and $\gamma_x\geq 0$ such that 

\begin{multline}\label{eq:sg-x}
\espp{\norm{G(\mbf{x})}_{\EL_p[t_0,t]}} \leq K_x |\mbf{x}(0)| +\\ \gamma_x \espp{\norm{(W(\mbf{e}),w)}_{\EL_p[t_0,t]}}.
\end{multline}

Based on \eqref{eq:sg-e} and \eqref{eq:sg-x}, we now proceed with small-gain arguments to complete the proof. For the remainder of the proof, we will omit the $\EL_p$-norm subscript to ease reading. That is,
\begin{align*}
	\espp{\norm{G(\mbf{x})}} &\leq 
 K_x|\mbf{x}(0)| + \gamma_x \espp{\norm{W(\mbf{e})}} + \gamma_x \norm{w} \\
	&\hspace{-1.8mm}\stackrel{\eqref{eq:sg-e}}{\leq} \gamma_x(K_e|\mbf{e}(0)|+\gamma_e\espp{\norm{G(\mbf{x})+Ew}}) \\
	&\qquad + K_x|\mbf{x}(0)| +  \gamma_x \norm{w}.
\end{align*}
The inequality above leads to
\begin{multline}\label{eq:sg-A}
	\espp{\norm{G(\mbf{x})}} \leq \frac{K_x|\mbf{x}(0)|+\gamma_x K_e|\mbf{e}(0)|}{1-\gamma_x\gamma_e}\\ + \frac{\gamma_x + \gamma_x\gamma_e|E|}{1 - \gamma_x\gamma_e} \norm{w}.
\end{multline}
Similarly, using \eqref{eq:sg-e} and \eqref{eq:sg-x}, we have that
\begin{align}\label{eq:sg-B}
	\espp{\norm{W(\mbf{e})}} &\leq \frac{K_e|\mbf{e}(0)|+\gamma_eK_x|\mbf{x}(0)|}{1 - \gamma_x\gamma_e} \nonumber \\
	&\hspace{1.5cm} + \frac{\gamma_x\gamma_e+\gamma_e|E|}{1-\gamma_x\gamma_e}\norm{w}.
\end{align}
To finalise the proof, we use \eqref{eq:sg-A} and \eqref{eq:sg-B} to obtain
\begin{align*}
	\espp{\norm{(G(\mbf{x}),W(\mbf{e}))}} &\leq \espp{\norm{G(\mbf{x})}} + \espp{\norm{W(\mbf{e})}} \\
	&\hspace{-2cm}\leq  \frac{K_x|\mbf{x}(0)|(1+\gamma_e) + K_e|\mbf{e}(0)|(1+\gamma_x)}{1-\gamma_x\gamma_e}  \\
	&\hspace{-0cm}\frac{\gamma_x + \gamma_e|E| + (1+|E|)\gamma_x\gamma_e}{1-\gamma_x\gamma_e}\norm{w} \\
	&\hspace{-2cm}\leq \frac{\max\{K_x(1+\gamma_e),K_e(1+\gamma_x)\}}{1-\gamma_x\gamma_e} (|\mbf{x}(0)|+|\mbf{e}(0)|)\\
	& + \tilde{\gamma} \norm{w}
\end{align*}
with $\tilde{\gamma}$ as per Theorem \ref{theo:Lp-overall}. Now, using the property $2xy\leq \epsilon x^2 + (1/\epsilon)y^2$ for any $x,y\in\R_{\geq 0}$ and $\epsilon>0$, we can bound
\begin{align*}
	|\mbf{x}(0)|+|\mbf{e}(0)| &= \sqrt{ |\mbf{x}(0)|^2 + |\mbf{e}(0)|^2 + 2|\mbf{x}(0)||\mbf{e}(0)| } \\
	&\leq \sqrt{ (1+\epsilon)|\mbf{x}(0)|^2 + (1+1/\epsilon)|\mbf{e}(0)|^2 } \\
	&\leq \left(\sqrt{\max\{1+\epsilon,1+1/\epsilon\}}\right) |(\mbf{x}(0),\mbf{e}(0))|.
\end{align*}
The proof is now complete.\qed

\subsection{Proof of Theorem \ref{theo:SS-explicit}}\label{sec:proof-SS-explicit}
The result follows from applying Corollary \ref{coro:SS-nolinear} to this specific scenario. Condition $(i)$ comes from Corollary \ref{coro:SS-nolinear}$(i)$, and from using the fact that the $\EL_2$ gain of the $\mbf{x}$--subsystem is given by $\gamma_x=\norm{P(j\omega)}_{\mathcal{H}_\infty}$, and also $K_x=|A_{21}|$, see e.g. \cite{khalil02}. The transmission rate condition $(ii)$ in the theorem's statement comes from Corollary \ref{coro:SS-nolinear}$(ii)$, but with the fact that $L=1/h$ in this setting. We show this in the following.
Particularly, we verify (for $p=2$) that $(iv)$ and $(v)$ of Corollary \ref{coro:SS-nolinear} hold for system $\mathcal{H}_1$ with $(A_{11},A_{21})$ detectable. We start with $(iv)$. Since the platoon adopts the SD protocol from Example \ref{ex:SD}, then $W(\mbf{e})=|\mbf{e}|$. Therefore, \eqref{eq:dWdt} for $\mathcal{H}_1$ reduces to
$ \left\langle \partial W/\partial \mbf{e},A_{21}\mbf{x}+A_{22}\mbf{e}+B_2w\right\rangle \leq |\dot{\mbf{e}}| 
    \leq |A_{22}||\mbf{e}|+|A_{21}\mbf{x}+B_2w|$.
Then, $L=|A_{22}|=1/h$, $\tilde{y}(\mbf{x},w)=A_{21}\mbf{x} + B_2w$, and thus $|E|=|B_2|=(1/h)\sqrt{1+k_d^2}$. These are, in fact, independet of $N$, as required by Corollary \ref{coro:SS-nolinear}. We now show that $(A_{11},A_{21})$ detectable implies $(v)$ in Corollary \ref{coro:SS-nolinear} is verified for $p=2$. Since the output of the $\mbf{e}$--subsystem is $W(\mbf{e})=|\mbf{e}|$, it suffices to show that $(A_{11},A_{21})$ detectable implies the $\mbf{x}$--subsystem is $\EL_2$ to $\EL_2$ detectable (in expectation) from $G(\mbf{x})$ to $\mbf{x}$, where $G(\mbf{x})=A_{21}\mbf{x}$.
Note that $(A_{11},A_{21})$ detectable implies there exists $L$ such that $A_{11}+LA_{21}$ is Hurwitz, and thus there exists $P=P^\top>0$ such that $(A_{11}+LA_{21})^\top P + P(A_{11}+LA_{21}) = -(\varepsilon/2) I$, for any $\varepsilon>0$. Let $U(\mbf{x})\coloneqq \mbf{x}^\top P \mbf{x}$. Then,
\allowdisplaybreaks
\begin{align*}
    &\left\langle \partial U/\partial \mbf{x},A_{11}\mbf{x}+A_{12}\mbf{e}+B_1w\right\rangle \\
    &\hspace{4mm}=\mbf{x}^\top \left[ (A_{11}+LA_{21})^\top P + P(A_{11}+LA_{21}) \right]\mbf{x} \\
    &\hspace{12mm}- \mbf{x}^\top \left[ A_{21}^\top L^\top P + P L A_{21} \right]\mbf{x} + \mbf{e}^\top A_{12}^\top P \mbf{x} \\
    &\hspace{12mm}+ w^\top B_1^\top P \mbf{x} +\mbf{x}^\top P A_{12}\mbf{e} + \mbf{x}^\top P B_1 w \\
    &\hspace{4mm}\leq -\frac{\varepsilon}{2} |\mbf{x}|^2 + 2|PL||A_{21}\mbf{x}|  |\mbf{x}| + 2|PA_{12}||\mbf{x}||\mbf{e}|
    \\
    &\hspace{12mm}+ 2|PB_1|\mbf{x}||w| \\
    &\hspace{4mm}\leq -\frac{\varepsilon}{4}|\mbf{x}|^2 - \frac{\varepsilon}{4}|\mbf{x}|^2 + \frac{3}{4}\varrho |\mbf{x}|^2 + \frac{4}{\rho} |PL|^2|A_{21}\mbf{x}|^2 \\
    &\hspace{12mm} +\frac{4}{\rho}|PA_{12}|^2|\mbf{e}|^2 
    +\frac{4}{\rho} |PB_1|^2|w|^2\\
    &\hspace{4mm}\leq -\frac{\varepsilon}{4}|\mbf{x}|^2 + \frac{4}{\rho} |PL|^2|A_{21}\mbf{x}|^2 \\
    &\hspace{12mm} +\frac{4}{\rho}|PA_{12}|^2|\mbf{e}|^2 
    +\frac{4}{\rho} |PB_1|^2|w|^2.
\end{align*}
where we used the property $2ab \leq (\varrho/4)a^2+ (4/\varrho)b^2$, and $\varepsilon$ satisfies $\varepsilon>3\varrho$. The last inequality implies the $\mbf{x}$--subsystem is $\EL_2$ to $\EL_2$ detectable in expectation from $A_{21}\mbf{x}$ to $\mbf{x}$, concluding the proof.\qed

\subsection{Proof of Theorem \ref{theo:SS-explicit-new}}\label{sec:proof-SS-explicit-2}
The proof follows identical steps to those of Theorem \ref{theo:SS-explicit}, albeit utilising different state-space matrices, given that the system is now characterised by $\mathcal{H}_2$  in \eqref{eq:HCS-2}. Essentially, we again apply Corollary \ref{coro:SS-nolinear}, but now to $\mathcal{H}_2$, and 
the main differences are that now $\gamma_x=\norm{Q(j\omega)}_{\mathcal{H}_\infty}$, $K_x=|\tilde{A}_{21}|$, $L=(1/h)\sqrt{1+k_d^2+k_p^2}$ and $|E|=|\tilde{B}_2|=|\tilde{C}B_1|=(1/h)\sqrt{1+k_d^2}$, which are independent of $N$.\qed


\bibliographystyle{ieeetr}
\bibliography{bibliography}


 





\end{document}